\documentclass[fleqn,10pt]{wlscirep}
\usepackage[utf8]{inputenc}
\usepackage[T1]{fontenc}

\usepackage[english]{babel}
\usepackage[superscript,biblabel]{cite}

\usepackage{amsmath}
\usepackage{graphicx}
\usepackage{subfigure}
\usepackage[colorlinks=true, allcolors=blue]{hyperref}
\usepackage{authblk}

\title{Coalitions in International Litigation: A Network Perspective}
\author[1,*]{R. Mastrandrea}
\author[2]{G. Antuofermo}
\author[3]{M. Ovadek}
\author[3]{T.Y.-C. Yeung}
\author[3]{A. Dyevre}
\author[4,5]{G. Caldarelli}
\affil[1]{IMT Alti Studi Lucca, Piazza S. Francesco 19, 55100 Lucca, Italy}
\affil[2] {Centro Studi Giuridici Francesco Carrara, Lucca, Italy}
\affil[3] {Centre for Empirical Jurisprudence, University of Leuven, Tiensestraat 45, 3000 Leuven, Belgium}
\affil[4]{DSMN Ca'Foscari, University of Venice, Via Torino 155, 30171 Venezia Mestre, Italy}
\affil[5]{ECLT Ca'Foscari, University of Venice, Dorsoduro 3246, 30123 Venice,  Italy}

\begin{abstract}
We apply network science principles to analyze the coalitions formed by European Union (EU) nations and institutions during litigation proceedings at the European Court of Justice. By constructing Friends and Foes networks, we explore their characteristics and dynamics through the application of cluster detection, motif analysis, and duplex analysis. Our findings demonstrate that the Friends and Foes networks exhibit disassortative behavior, highlighting the inclination of nodes to connect with dissimilar nodes. Furthermore, there is a correlation among centrality measures, indicating that member states and institutions with a larger number of connections play a prominent role in bridging the network. An examination of the modularity of the networks reveals that coalitions tend to align along regional and institutional lines, rather than national government divisions. Additionally, an analysis of triadic binary motifs uncovers a greater level of reciprocity within the Foes network compared to the Friends network.
\end{abstract}

\affil[*]{Rossana.Mastrandrea@imtlucca.it}
\begin{document}

\flushbottom
\maketitle
%
%
\thispagestyle{empty}
\section*{Introduction}
\medskip
The digital revolution\cite{editorial2008flood} has ushered in an era of unprecedented access to vast amounts of data, revolutionizing social-scientific research and opening up new avenues for quantitative approaches.\cite{natcomed2018data} Network science, a powerful tool for representing and visualizing complex systems \cite{caldarelli2007scale}, has been increasingly applied across various domains, from the study of scientific advancements\cite{gates2019nature} to finance\cite{bardoscia2021physics} and social systems\cite{lazer2009social}.

Although the application of network science to social behaviour and human institutions predates the digital revolution as illustrated by Jonathan L. Moreno's famous sociograms\cite{moreno1934shall} and pioneering studies on self-organised segregation phenomena\cite{schelling1969models} and social cooperation \cite{axelrod1981evolution}, law has long been perceived as a field beyond the reach of quantitative modelling. However, with digitalisation making considerable progress, the legal field has also witnessed growing interest in network science methods to model case citation dynamics\cite{fowler2007network,lupu2012precedent,larsson2017speaking,lettieri2016computational,sadl2017can,mones2021emergence}, the evolution and structure of legislation \cite{katz2020complex,koniaris2018network} as well as professional networks of judges \cite{katz2010hustle} and law professors \cite{ruhl2015measuring}. While untangling the complexity of normative structures\cite{boulet2019law,koniaris2018network}, this literature has delivered new insights on the operations of legal institutions\cite{kosack2018functional} and the possibility to predict case citations \cite{mones2021emergence}.

 In this context, we present a network analysis of coalitions in litigation before the Court of Justice of the European Union (CJEU). The CJEU adjudicates disputes over the legality of EU acts, the interpretation of EU treaties and legislation and state compliance with EU policies. Due to the impact and implications of its rulings across the bloc, the CJEU is regarded as one of the world's most powerful judicial bodies. National governments and EU institutions may appear before the Court as complainant or as defendant. They may also intervene in proceedings to express support or opposition to the EU member state or the EU institution party to the case. 
 
 We use network analysis to study the coalition patterns emerging from this data. We construct networks to model the web of directed "friendly" and "unfriendly" connections between intervening states and parties. We examine centrality, modularity and triadic motifs both in the "Friends" and "Foes" networks over time. Additionally, we conduct a multiplex analysis and merge the two networks to gain further insights.

We highlight the following findings. Firstly, Friends and Foes networks (see below) display a disassortative behaviour, implying a tendency for nodes to connect with dissimilar nodes rather than similar ones. Secondly, strong correlations among centrality measures suggest that member states and institutions with a higher number of connections simultaneously play a prominent role in bridging the nodes. Thirdly, the modularity of networks points to alignments along regional lines and divisions between EU institutions and member states consistently with results from social science research on European integration. Finally, we find a greater degree of reciprocity within the Foes network compared to the Friends network, suggesting a higher level of mutual opposition and conflict among nodes in the Foes network.

\section{Materials and Methods}
\subsection{Data}
\medskip
Our data set consists of 625 cases filed with the CJEU between 1977 and 2018. We use web-scraping methods to identify and extract cases with at least one third party intervention from the EUR-Lex website (\url{www.eur-lex.eu}). Cases in our data set are either initiated by a member state against an EU institution (annulment action) or initiated by the European Commission against a member state (infringement action). We only consider cases involving coalitions, i.e. cases which feature at least one intervention or in which two states or more act as plaintiff/defendant. While governments are also allowed to present observations in cases passed on to the CJEU by national courts (so-called ?preliminary rulings? in EU law jargon), these cases only address matters of interpretation and do not determine the final outcome of the legal dispute at hand. The observations presented by intervening governments in preliminary rulings are ambiguous and do not clearly indicate which side they are meant to support. For these reasons, they are excluded from our analysis. 

The number of third party interventions varies between 1 and 20 per case. In this data, 29 cases do not feature interventions but are also included as they feature two or more countries/institutions as main parties on the same side of the dispute. While governments have enjoyed the right to intervene in CJEU proceedings since the inception of European integration in the 1950s, the first intervention occured in 1977. To explore the evolution of coalition dynamics over time, we divided the data in eight periods: 1977-1981, 1982-1986, 1987-1991, 1992-1996, 1997-2001, 2002-2006, 2007-2011 and 2012-2018.

\subsection{Network Structure}
\medskip
We constructed two directed, weighted networks. The first is a same-side network, which we refer to as \textit{Friends} network; the second an opposite-side network, which we refer to as \textit{Foes} network. In our two networks, a {\bf Node} represents a country or an EU institution involved in a case either as plaintiff/defendant or as intervening third party. An {\bf Edge} is drawn between two nodes if they are on the same (Friends network) or opposite (Foes network) side of the case, whereby the intervening party is the source while the plaintiff/defendant and their co-interveners are the target. We make two exceptions to this rule. First, we omit to draw an edge between the Commission and the main party to the case in infringement proceedings initiated by the European Commission. Similarly, we do not draw any edge between the case initiator and the Council of the European Union in annulment proceedings. Because infringement actions are brought by the Commission and all annulment cases are directed against legislation approved by the Council, these edges would merely reflect the proportion of infringement and annulment cases in our data.
For the purpose of investigating coalition dynamics, these edges are, therefore, uninformative. 


Edges are weighted according to the number of cases involving the two nodes as friend/foe in the corresponding period.
We then define two adjacency matrices of size $N$: our $fr$ and $fo$ matrices.  They represent the weighted adjacency matrices associated to, respectively, the Friends and Foes network and we indicate the element of such matrices with $fr_{ij}$ and $fo_{ij}$. {\em i.e.}
\begin{equation}
\left[
W^{fr} \equiv (w^{fr}_{ij})_{1 \le i,j\le N} \quad \quad W^{fo} \equiv (w^{fo}_{ij})_{1 \le i,j\le N}
\right]
\end{equation}

\noindent 

\subsection{Node Importance and Network Organisation}
\medskip
Node importance is assessed using various centrality measures. Some of those centrality measures are for example designed to capture the ability of a node to spur or stop epidemic processes within a network, others to measure the bridging value of a connection and therefore we lack a unique all-purpose centrality measure. For this reason, the importance of a given centrality metric is often a matter of the context. In our study, we focus on four measures of node centrality that highlight different aspects of our system: (i) degree centrality, which represents a simple (binary), local measure considering only the first order connections of a node; (ii) strength centrality. which as a natural extension of degree centrality, includes the effect of edge-values on node importance; (iii) betweenness centrality, which is defined as a higher-order measure of node importance taking into account the shortest paths connecting any two nodes in the network, thus focusing on its possible key role in diffusion processes; and (iv) page-rank centrality \cite{page1998pagerank}, which functions as a global measure recursively taking into account the centrality of all the node's neighbours -- in other words, it assigns a measure of importance to a node according to the centrality of its neighbours.


We examine triangular closure and compare motif occurrence to a null model. To detect community structures, we apply the Louvain algorithm. Our multiplex analysis models the two networks as layers of a duplex, with the analysis focusing on the overlapping node degree/strength (sum of node degree/strength over the two layers), the multiplex participation coefficient with respect to a chosen local quantity and the z-score of the overlapping degree/strength to identify - if any - peculiar nodes\cite{avalos2018emergent}.

Finally, we merged the two networks into a single network by computing the frequency of relationships for each pair of nodes: $w^{fr}_{ij}/(w^{fr}_{ij} + w^{fo}_{ij})$.
We consider this network representation complementary to the multiplex perspective -- offering a more intuitive visualization of the relationships occurring among countries/institutions and allowing the study of additional topological properties, although at the price of losing some features of the two layers.

\section{Results}
\medskip
We first examine basic topological local properties (size, volume, degree, strength) before turning to centrality measures and community structures.

\subsection{Basic Network Properties}
\begin{figure}[!ht]
\centering
{\includegraphics[width=0.9\textwidth]{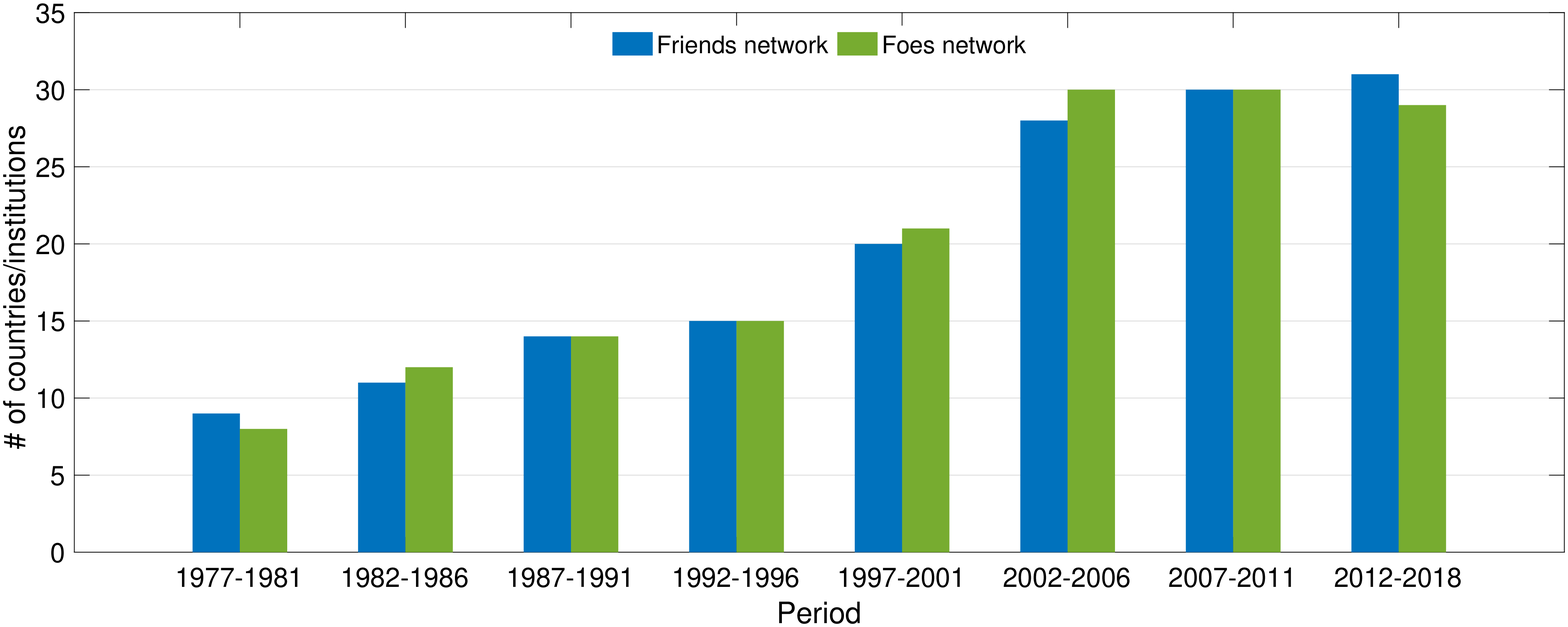}}
{\includegraphics[width=0.9\textwidth]{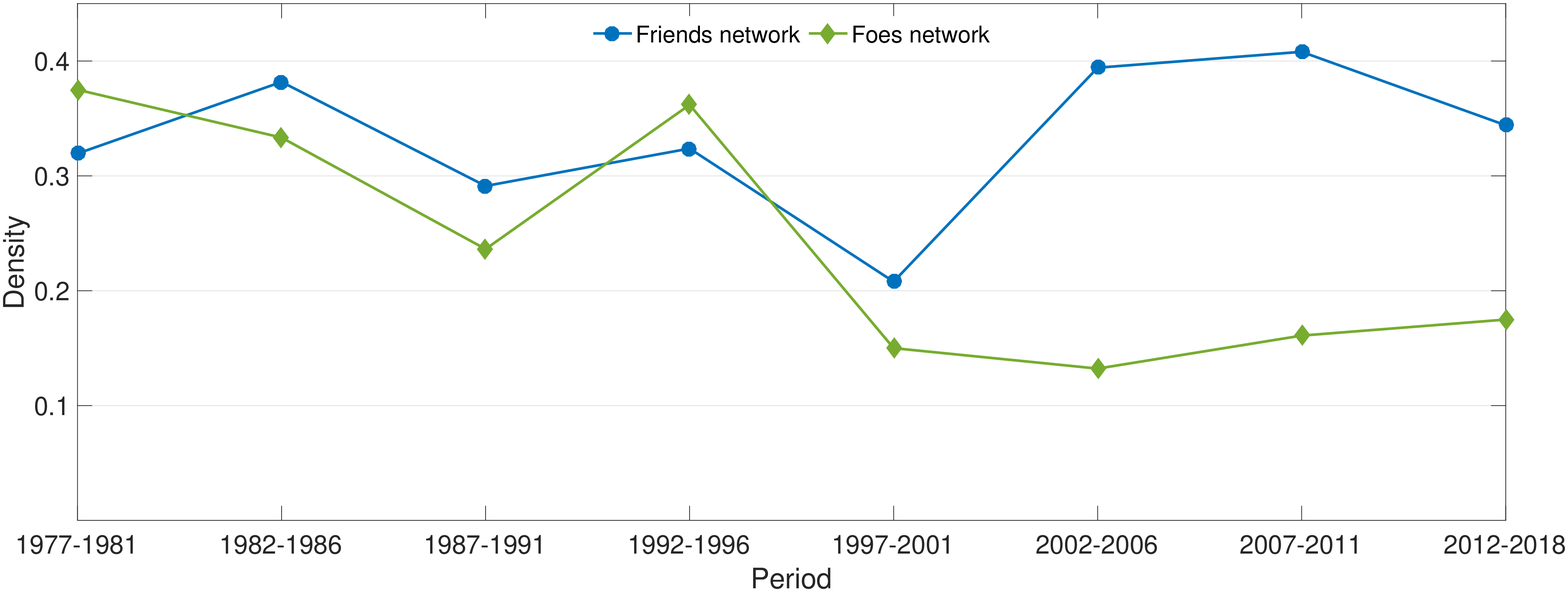}}
{\includegraphics[width=0.9\textwidth]{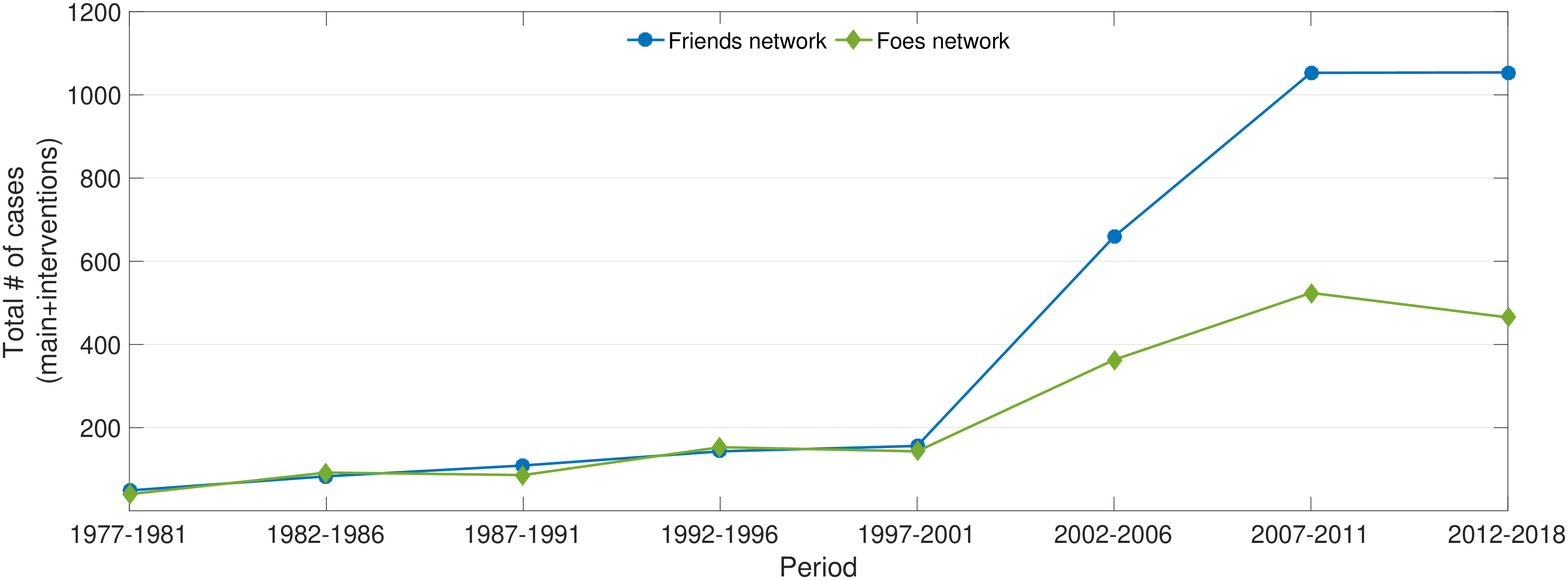}}
\caption{{\bf Network properties.} Size (top); density (middle); total weight/volume (bottom).\label{fig:size}}
\end{figure}

\medskip
Our entire data set features 36 countries and institutions (see Table in Appendix). Yet the network size never exceeds 31 nodes for any of the eight periods, as shown in Figure \ref{fig:size}. Successive increases in network size reflect the impact of enlargement (19 countries joined the EU between 1981 and 2013), treaty revisions (which created new institutions, such as the European Central Bank) and more frequent litigation, which has provided governments and EU institutions with more opportunities to intervene in judicial proceedings. Whereas \textit{Friends} and \textit{Foes} networks increase at the same pace (Fig.\ref{fig:size}, top), the density of the Friends network appears higher in the last 3 periods after starting from similar values (Fig.\ref{fig:size},  middle). The total number of cases rises in tandem with the number of agents involved, but with greater speed for the Friends network (Fig.\ref{fig:size}, bottom). 

\begin{figure}[!ht]
\centering
{\includegraphics[width=1\textwidth]{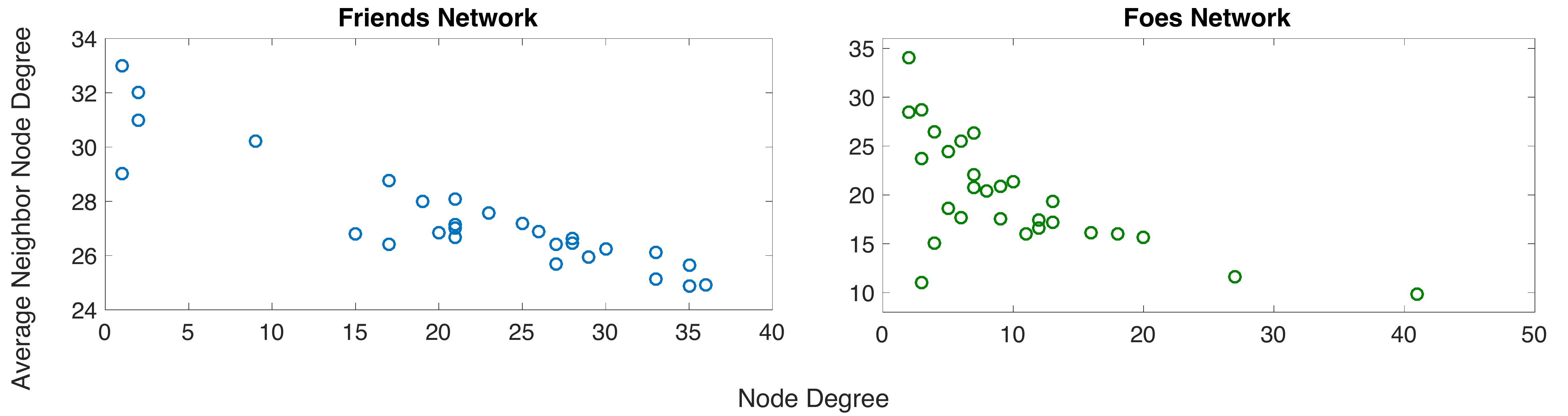}}
\caption{{\bf Disassortativity} Total node degree versus the Average Neighbor Node Degree for Friends and Foes Network in the period 2012-2018. \label{fig:dis}}
\end{figure}

Both Friends and Foes networks exhibit binary disassortativity when modelled as binary, although not when modelled as weighted. Figure \ref{fig:dis} shows the scatter plot for the binary model for the last period in our data set, 2012-2018. Disassortativity suggests a tendency for countries and institutions involved in a high number of lawsuits to be connected with countries and institutions less active in the litigation process. This property is a manifestation of structural disparities in the involvement of institutions and member states in EU-level legal disputes. Some institutions (e.g. ECB) have authority over narrow policy domains, limiting the range of cases in which they may be involved, while member states differ considerably in economic size, political influence and familiarity with EU law litigation \cite{ovadek2021supranationalism,kaeding2005mapping,mattila2009roll,thomson2009actor}. 

\subsection{Node Centrality}
\medskip

\begin{figure}[!ht]
\centering
\subfigure[In degree centrality]
{\includegraphics[width=0.45\textwidth]{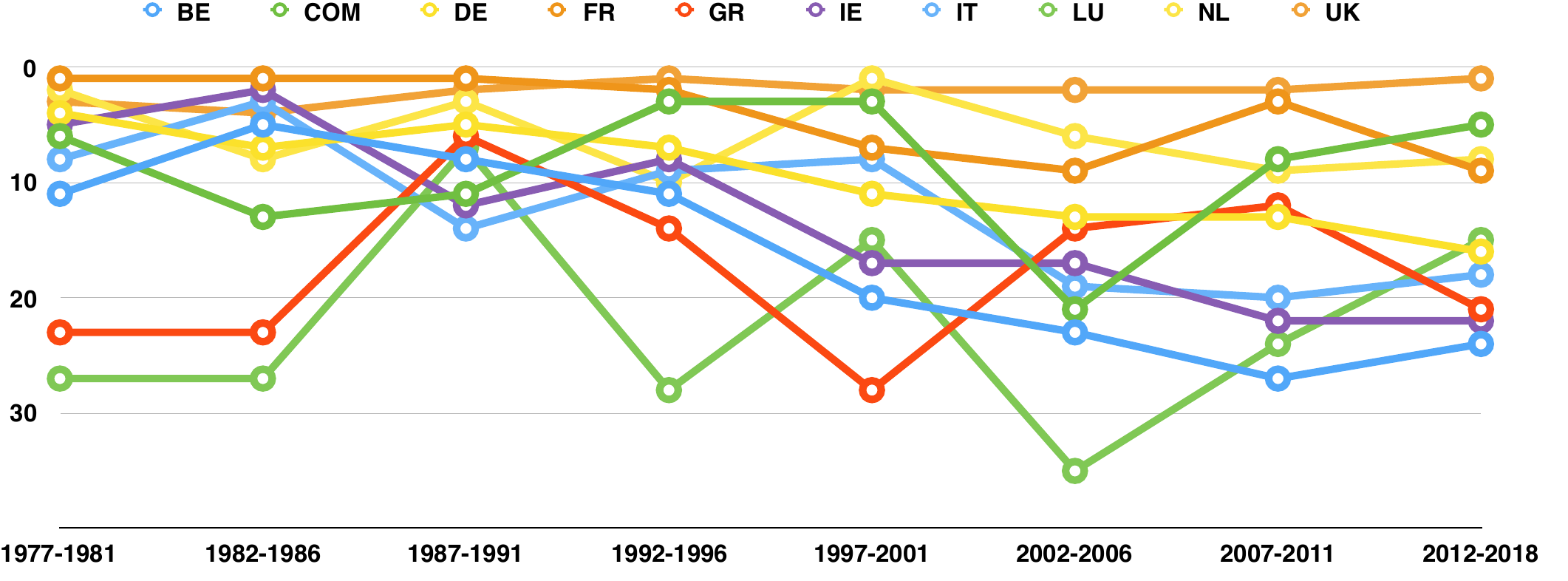}}
\hspace{5mm}
\subfigure[Out degree centrality]
{\includegraphics[width=0.45\textwidth]{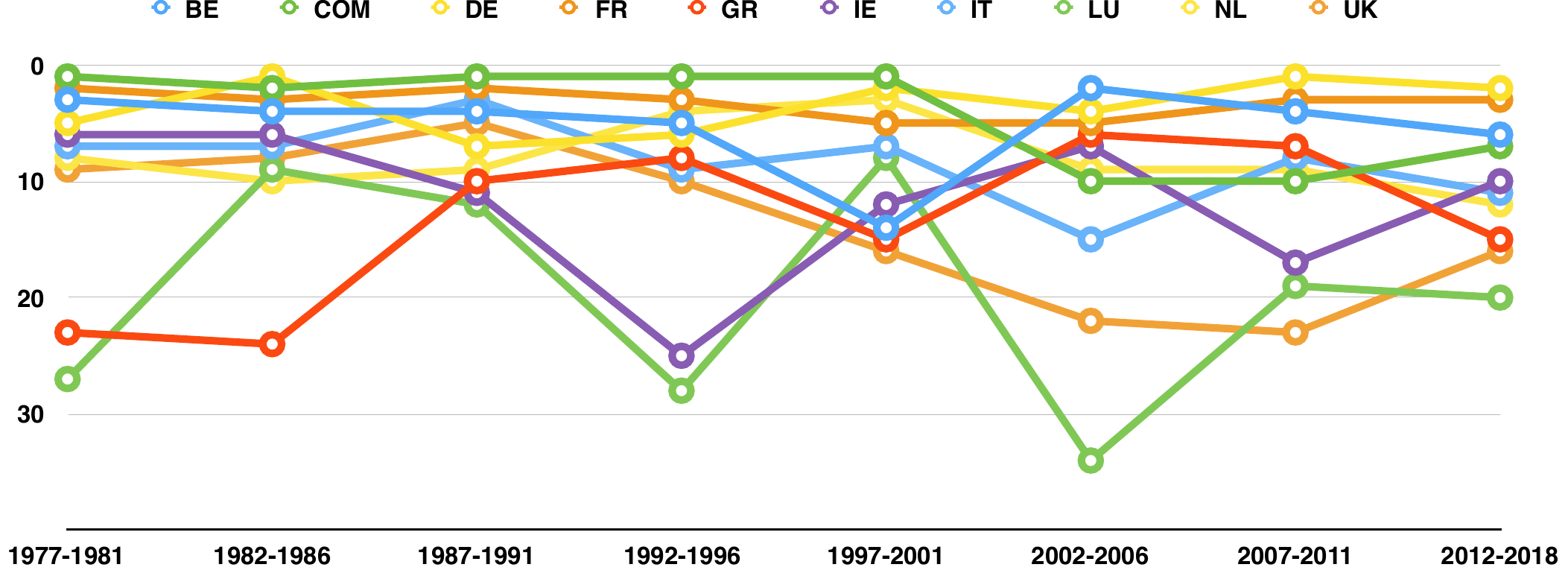}}
\subfigure[Betweenness centrality]
{\includegraphics[width=0.45\textwidth]{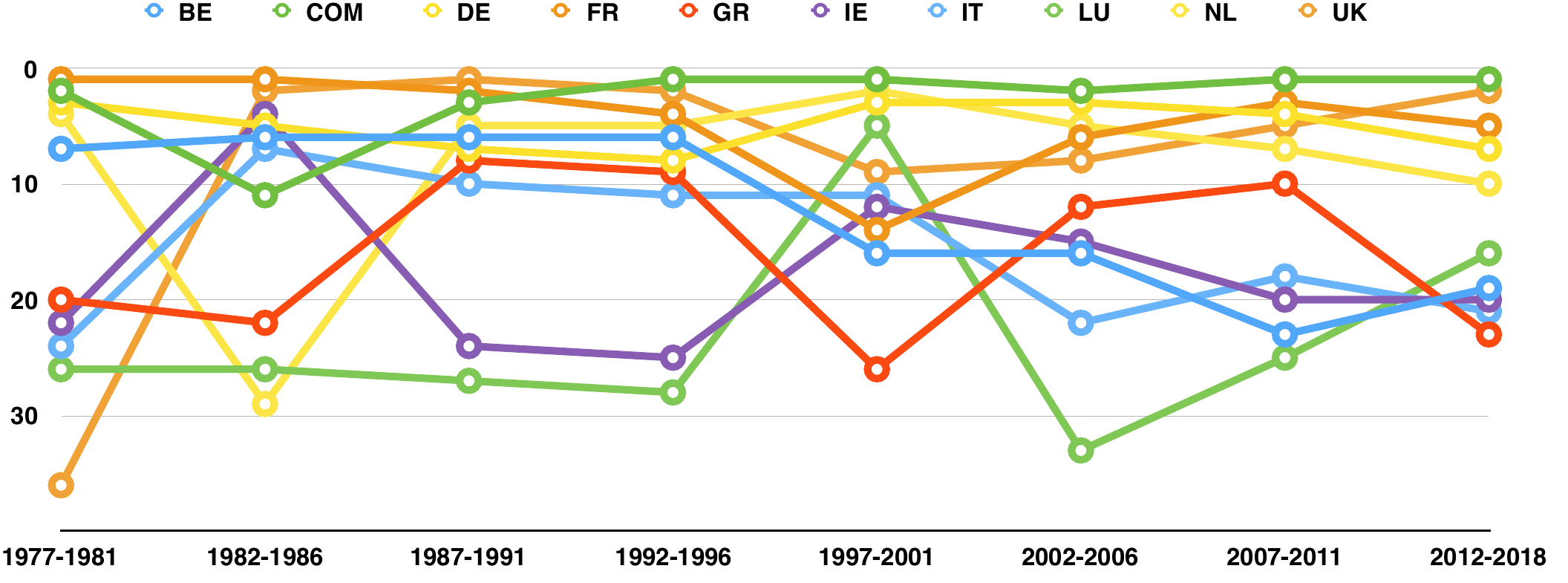}}
\hspace{5mm}
\subfigure[Page rank centrality]
{\includegraphics[width=0.45\textwidth]{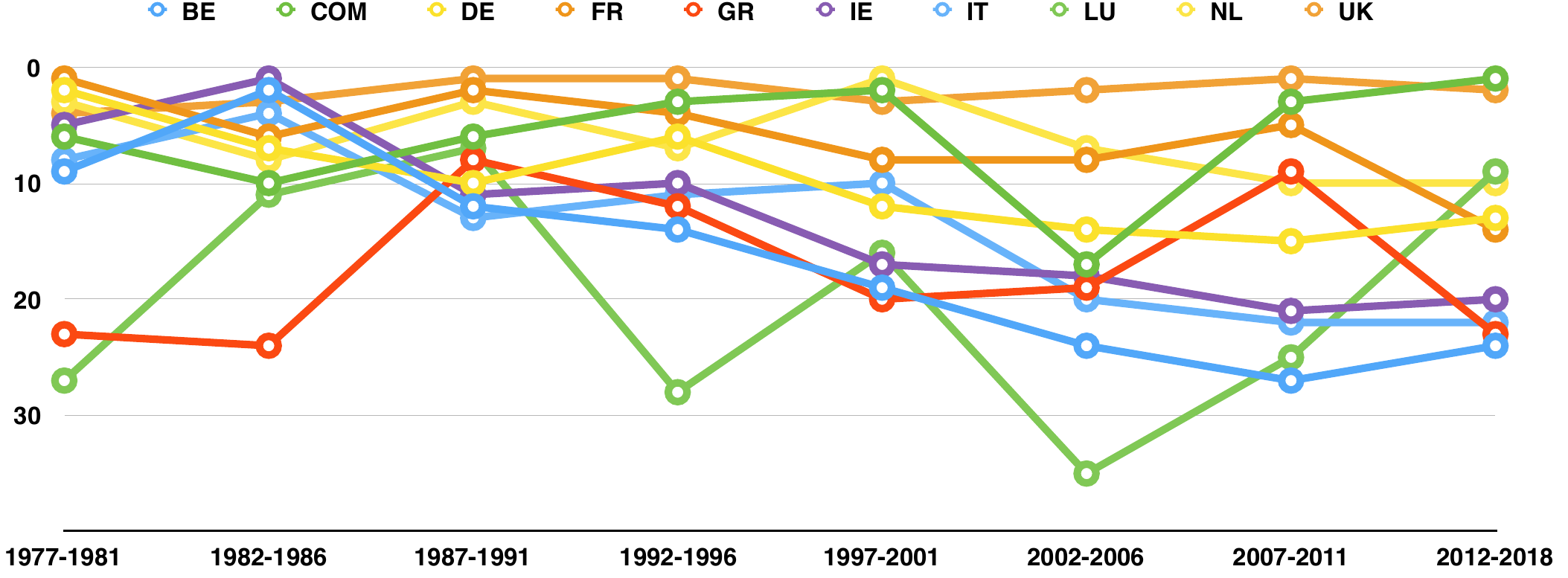}}
\caption{{\bf Centrality ranking for Friends network} Only ten countries/institutions are included in the figures as the most involved in both Foes and Friends networks in the whole period (BE = Belgium; COM = European Commission;  DE = Germany; FR = France; GR = Greece; IE = Ireland; IT = Italy; LU = Luxembourg; NL = Netherlands; UK = United Kingdom.)
\label{fig:cenFr}}
\end{figure}

\begin{figure}[!ht]
\centering
\subfigure[In degree centrality]
{\includegraphics[width=0.45\textwidth]{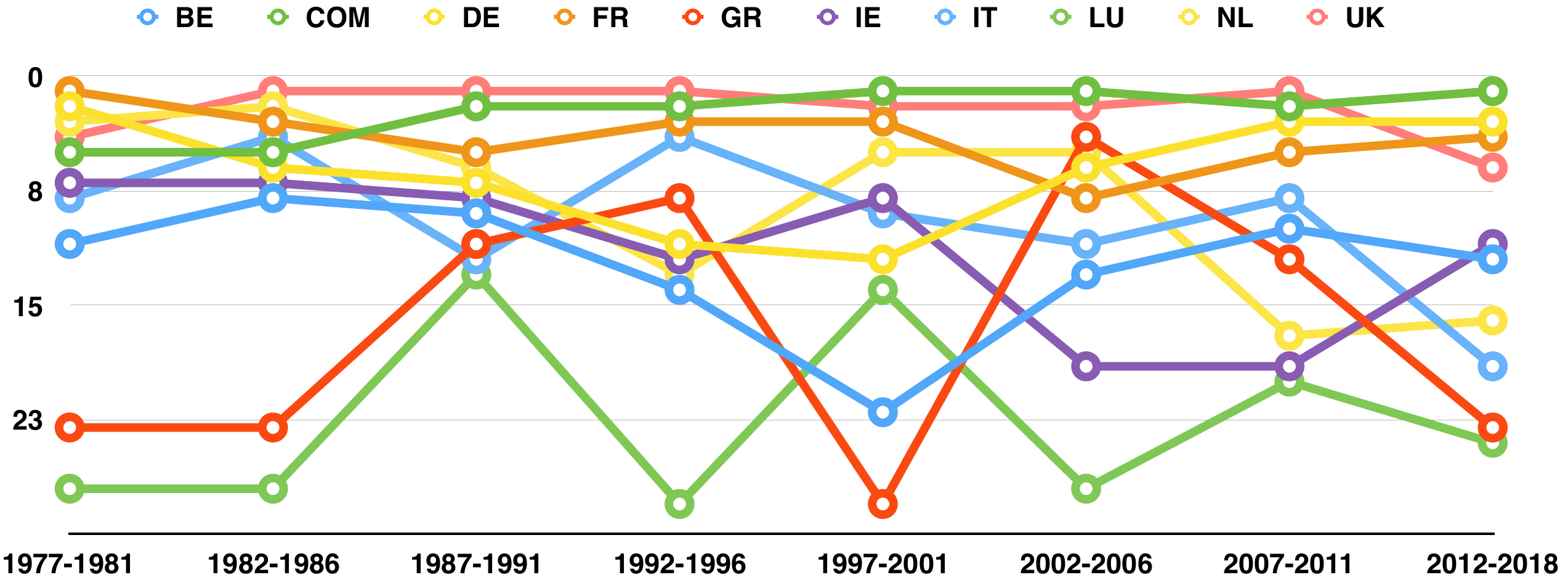}}
\hspace{5mm}
\subfigure[Out degree centrality]
{\includegraphics[width=0.45\textwidth]{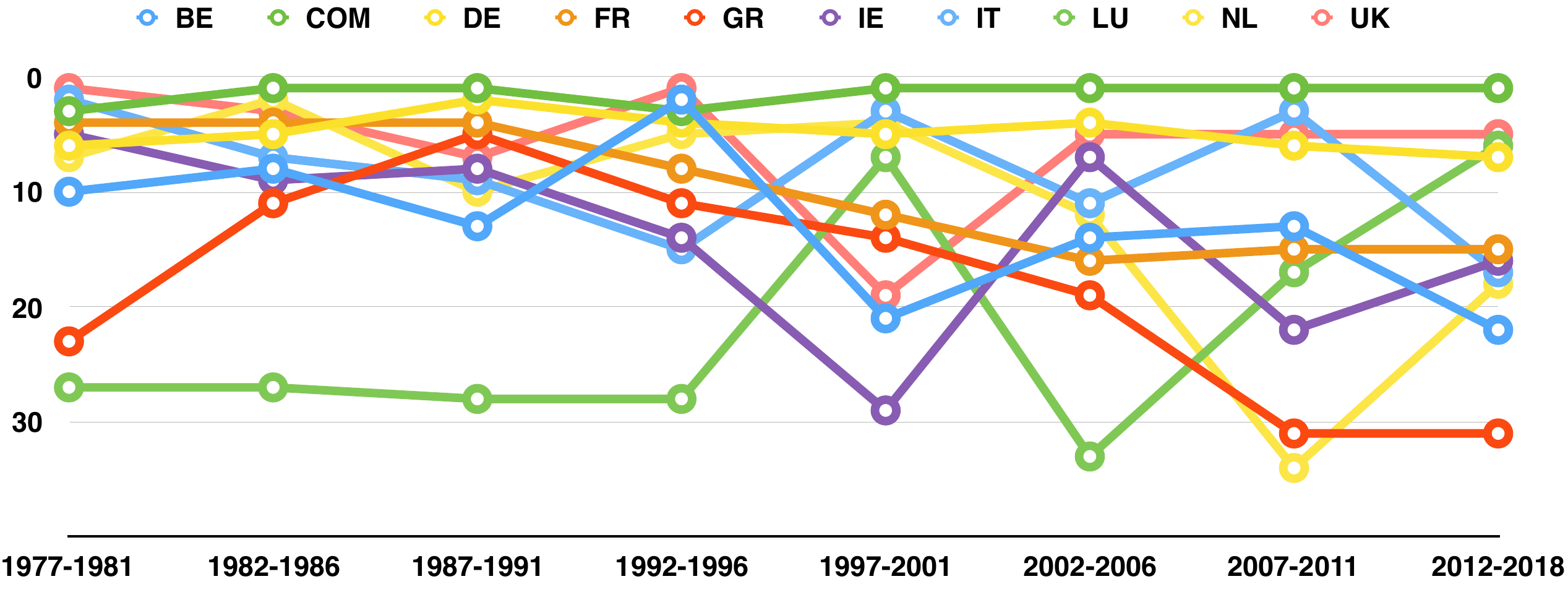}}
\subfigure[Betweenness centrality]
{\includegraphics[width=0.45\textwidth]{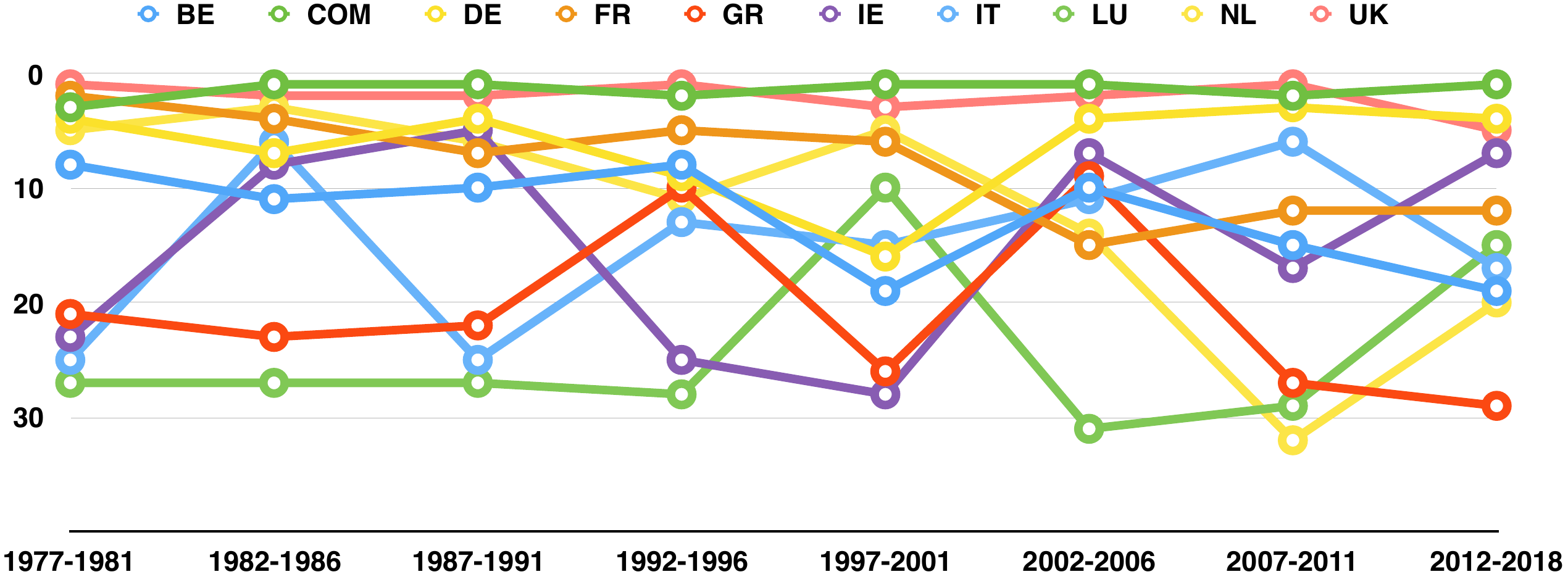}}
\hspace{5mm}
\subfigure[Page rank centrality]
{\includegraphics[width=0.45\textwidth]{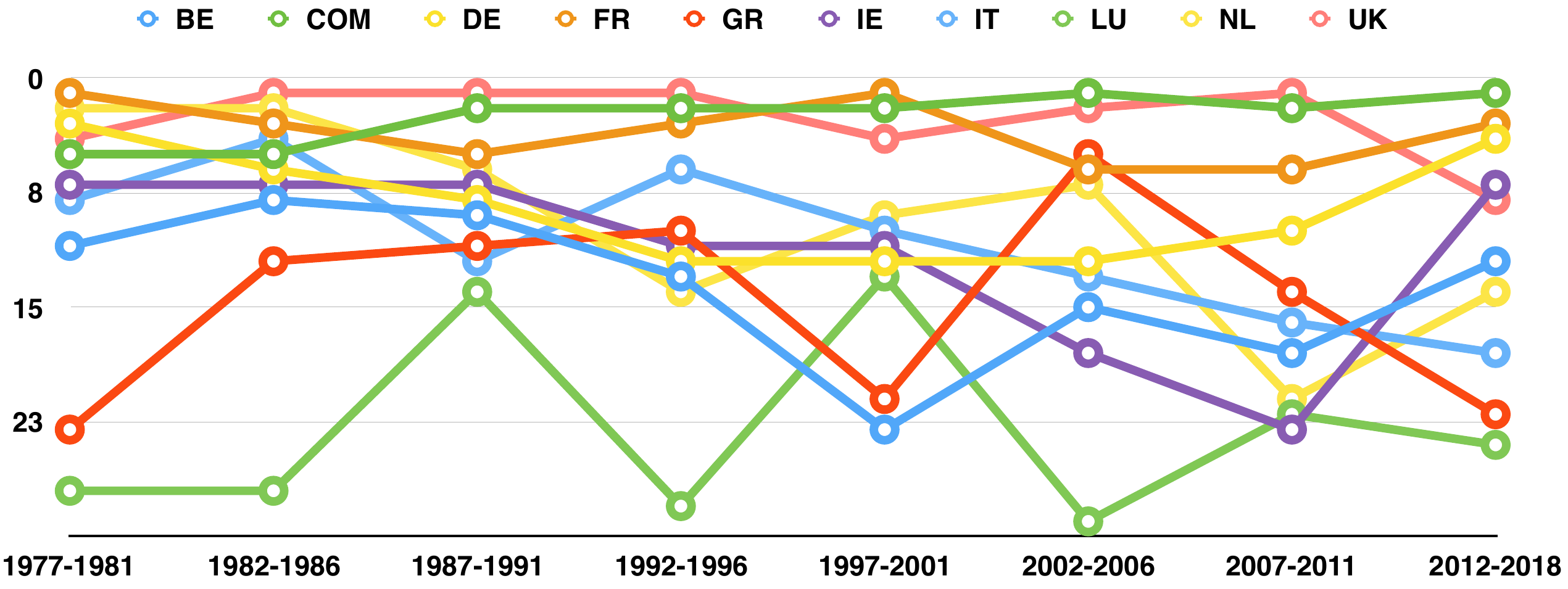}}
\caption{{\bf Centrality ranking for Foes network} Only ten countries/institutions are included in the figures as the most involved in both Foes and Friends networks in the whole period (BE = Belgium; COM = European Commission;  DE = Germany; FR = France; GR = Greece; IE = Ireland; IT = Italy; LU = Luxembourg; NL = Netherlands; UK = United Kingdom.)
\label{fig:cenFo}}
\end{figure}

Figure \ref{fig:cenFr} and \ref{fig:cenFo} report node rankings according to our four centrality measures over the eight periods for both Friends and Foes. Here the analysis is restricted to the countries and institutions appearing in all periods in both networks to ensure meaningful temporal comparisons.  
France dominates the Friends rankings in the first three periods, whereas later periods are dominated by the UK, the Czech Republic, Finland and the European Commission. For the Foes network, centrality rankings are dominated by the UK and the European Commission. The fact that the UK and the Commission score high on out-degree centrality as well as on the other three measures indicate that they are both initiators and targets of hostile interventions.

\begin{figure}[!ht]
\centering
\subfigure[Friends network]
{\includegraphics[width=0.45\textwidth]{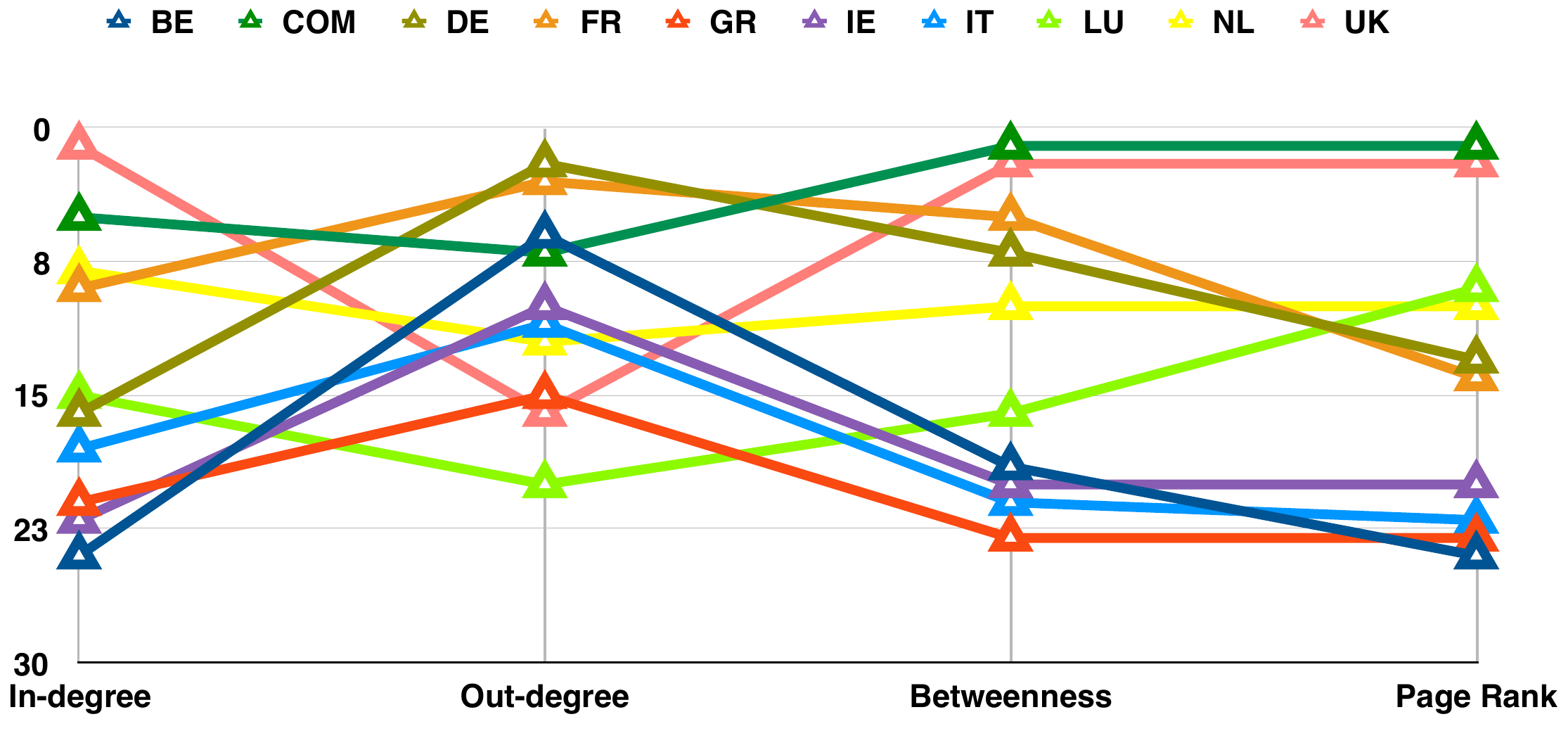}}
\hspace{5mm}
\subfigure[Foes network]
{\includegraphics[width=0.45\textwidth]{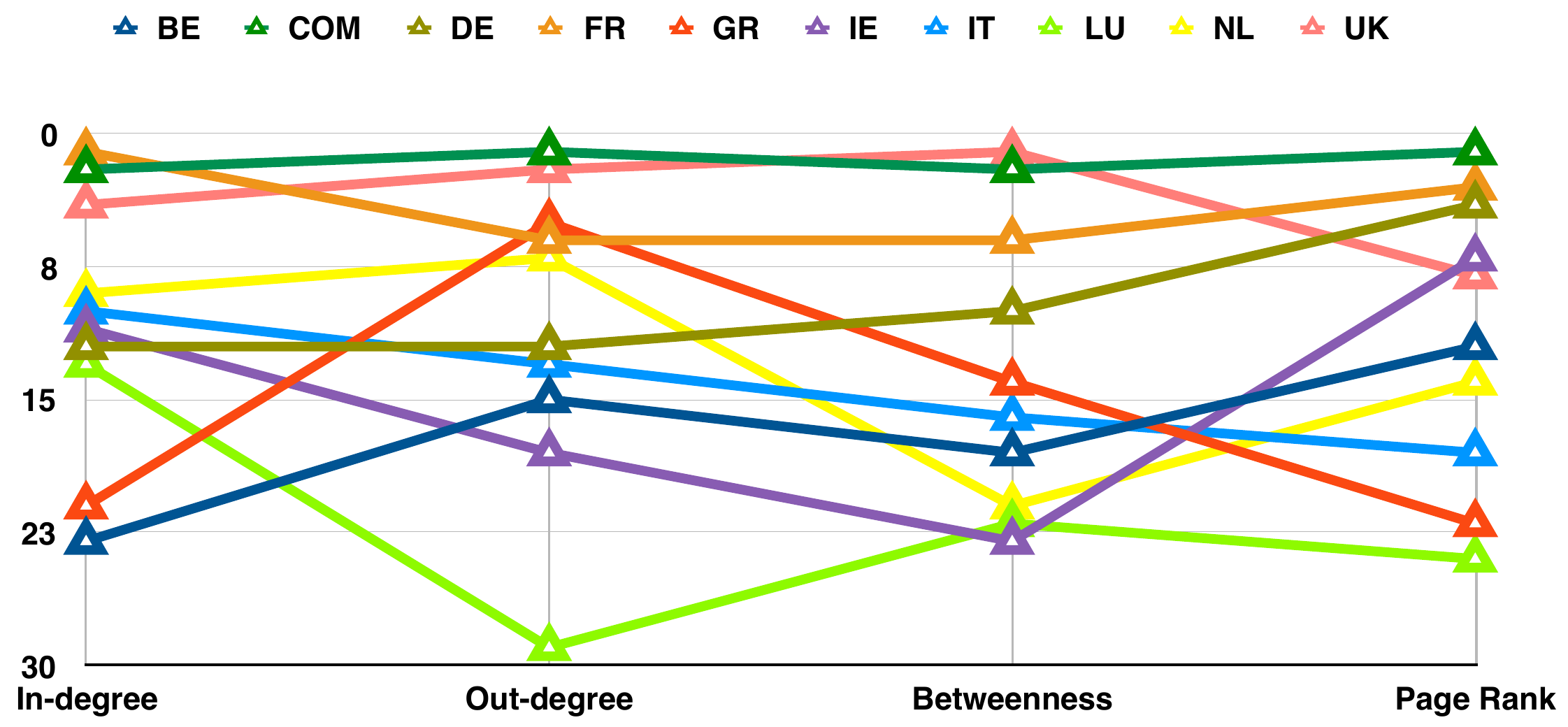}}

\caption{{\bf Centrality ranking comparison} Ranking for the four centrality measures in the last period under study for (a) Friend network and (b) Foes network.Only ten countries/institutions are included in the figures as the most involved in both Foes and Friends networks in the whole period (BE = Belgium; COM = European Commission;  DE = Germany; FR = France; GR = Greece; IE = Ireland; IT = Italy; LU = Luxembourg; NL = Netherlands; UK = United Kingdom.) 
\label{fig:cenAll}}
\end{figure}

Figure \ref{fig:cenAll} report node rankings for the period 2012-2018. Ranking for the Commission differ little across centrality in both the Friends and Foes networks. In the Friends network Germany ranks highest on out-degree centrality, indicating frequent friendly interventions. The UK's in-degree score in the Friends network and out-degree score in the Foes network reveal active intervention both against and in support of other EU actors.

\begin{figure}[!ht]
\centering
{\includegraphics[width=0.45\textwidth]{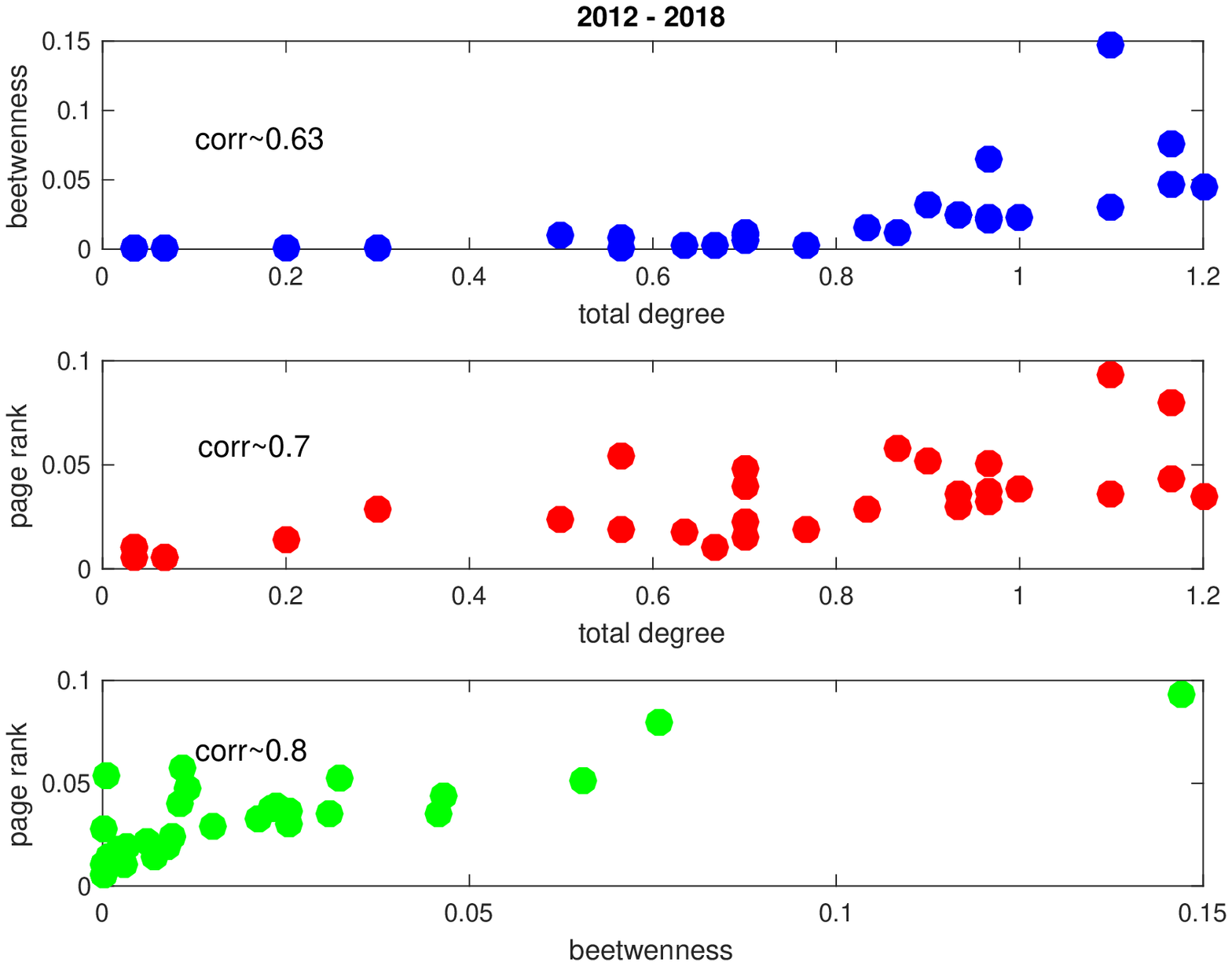}}
\hspace{5mm}
{\includegraphics[width=0.45\textwidth]{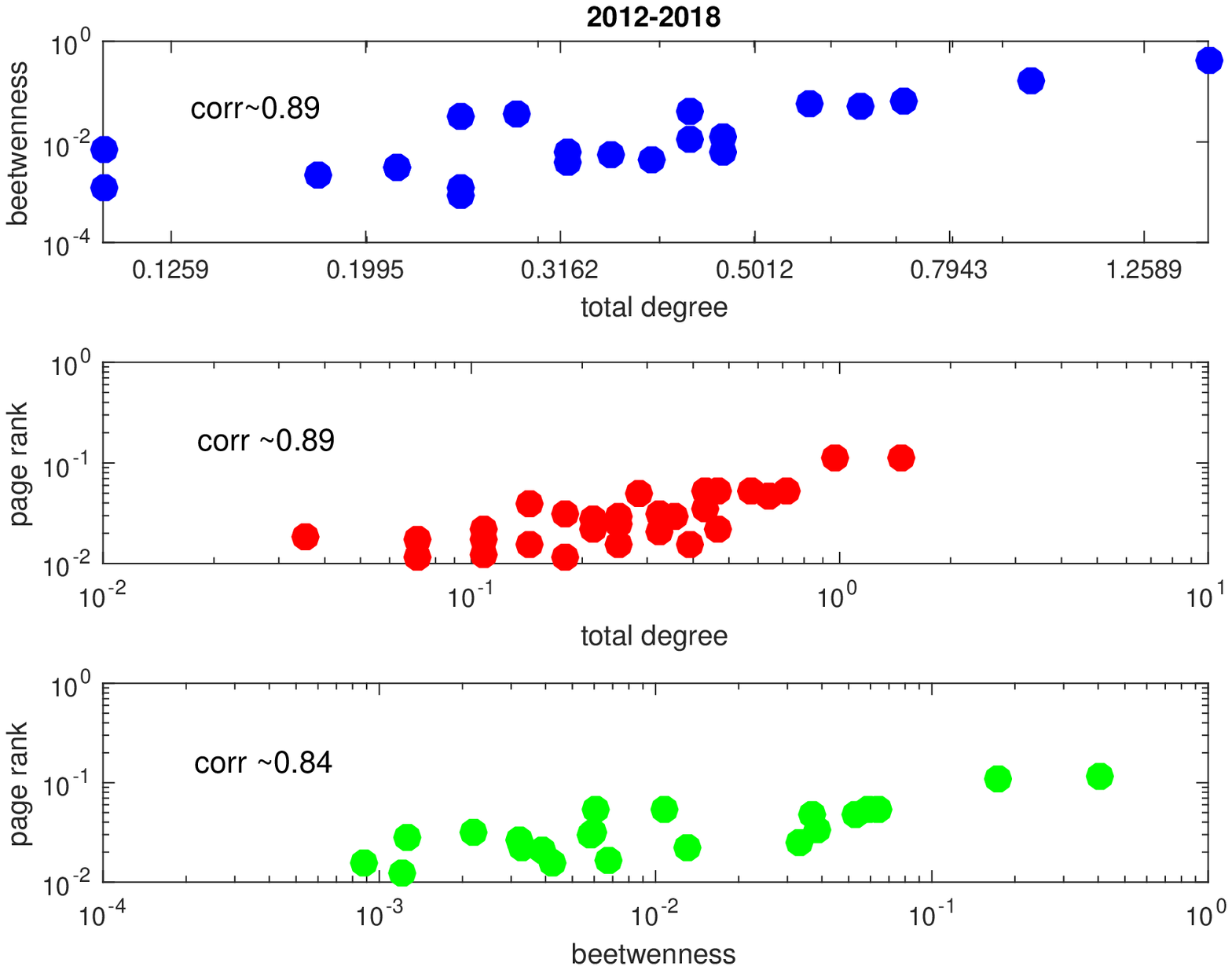}}
\caption{{\bf Correlation centrality measures.} Friends (left); Foes (right).\label{fig:corr}}
\end{figure}

The scatter plots in figure \ref{fig:corr} display correlation values among the four centrality measures for the last period. We observe high correlations between all centrality measures (for the sake of simplicity, we report total degree rather than in and out-degree). High correlation values are not a necessary property of every network. The central node of a star network, for instance, will possess both high degree and high betweennes centrality, whereas a node connecting the central node of two star networks will exhibit low degree but high betweenness centrality. Correlation values thus indicate the extent to which nodes tend to play the same role in the network. More specifically, in our networks, the degree-betweenness correlation highlights the fact that countries/institutions with a great number of neighbours also a fundamental role in bridging the network.


 \clearpage

\subsection{Motifs} 
\medskip

\begin{figure}[!ht]
\centering
{\includegraphics[width=0.3\textwidth]{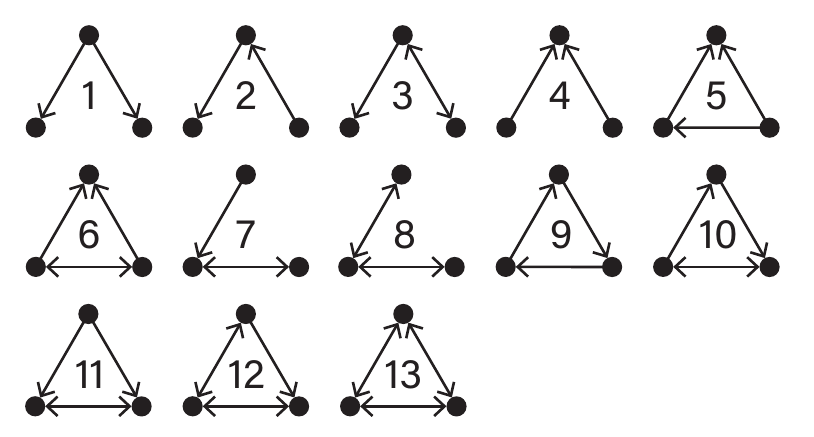}}
\hspace{50mm}
{\includegraphics[width=0.65\textwidth]{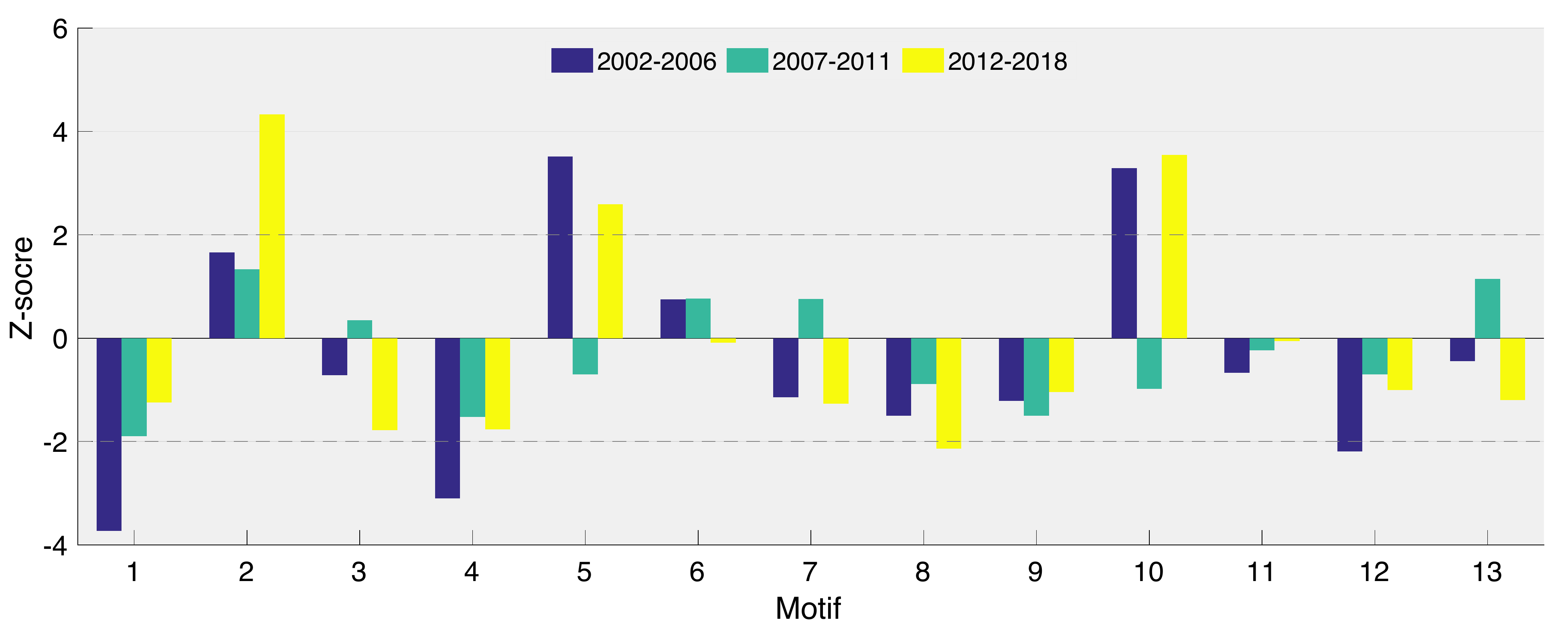}}
{\includegraphics[width=0.65\textwidth]{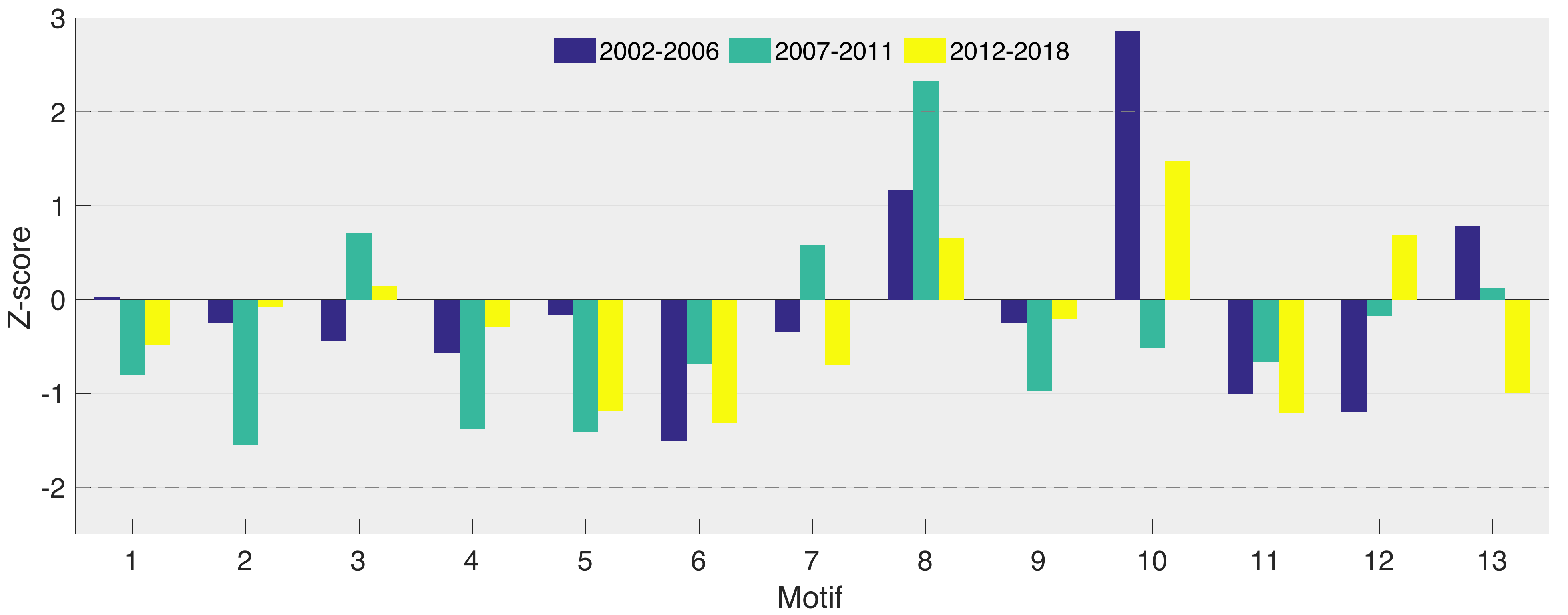}}
\caption{{\bf Binary motifs.} Friends network (top) and Foes network (bottom) for the last three periods analyzed.\label{fig:mot}}
\end{figure}

To detect patterns of coalition formation, we performed an analysis of recurrent triadic binary motifs (Fig. \ref{fig:mot}, top), comparing their occurrences in our networks with null models sharing the same degree distribution (see Methods for details). We compare the three periods 2002-2006, 2007-2011 and 2012-2018, which are sufficiently comparable in terms of network size so as not to affect the z-score. 

Figure \ref{fig:mot} shows the z-scores for each motif in the three periods. As regards the Friends network, motif 1 and 4 appear to be overestimated by the null model, open triangles with either two exiting or two entering links are less frequent than expected for the period 2002-2006. Motif 5 and 10, by contrast, are more frequent in the Friends network than in the null models for two periods 2002-2006 and 2012-2018. The frequency of motif 10 in these two periods suggests limited reciprocation in friendly support. Interestingly, reciprocation seems more pronounced in the Foes network. Motif 8 and 10, both triangles with reciprocated links, are abundant, especially in the period 2007-2011 (motif 8) and 2002-2006. These patterns suggest that member states and institutions tend to reciprocate hostile interventions.

\subsection{Communities}
\begin{figure}[!ht]
\centering
{\includegraphics[width=0.45\textwidth]{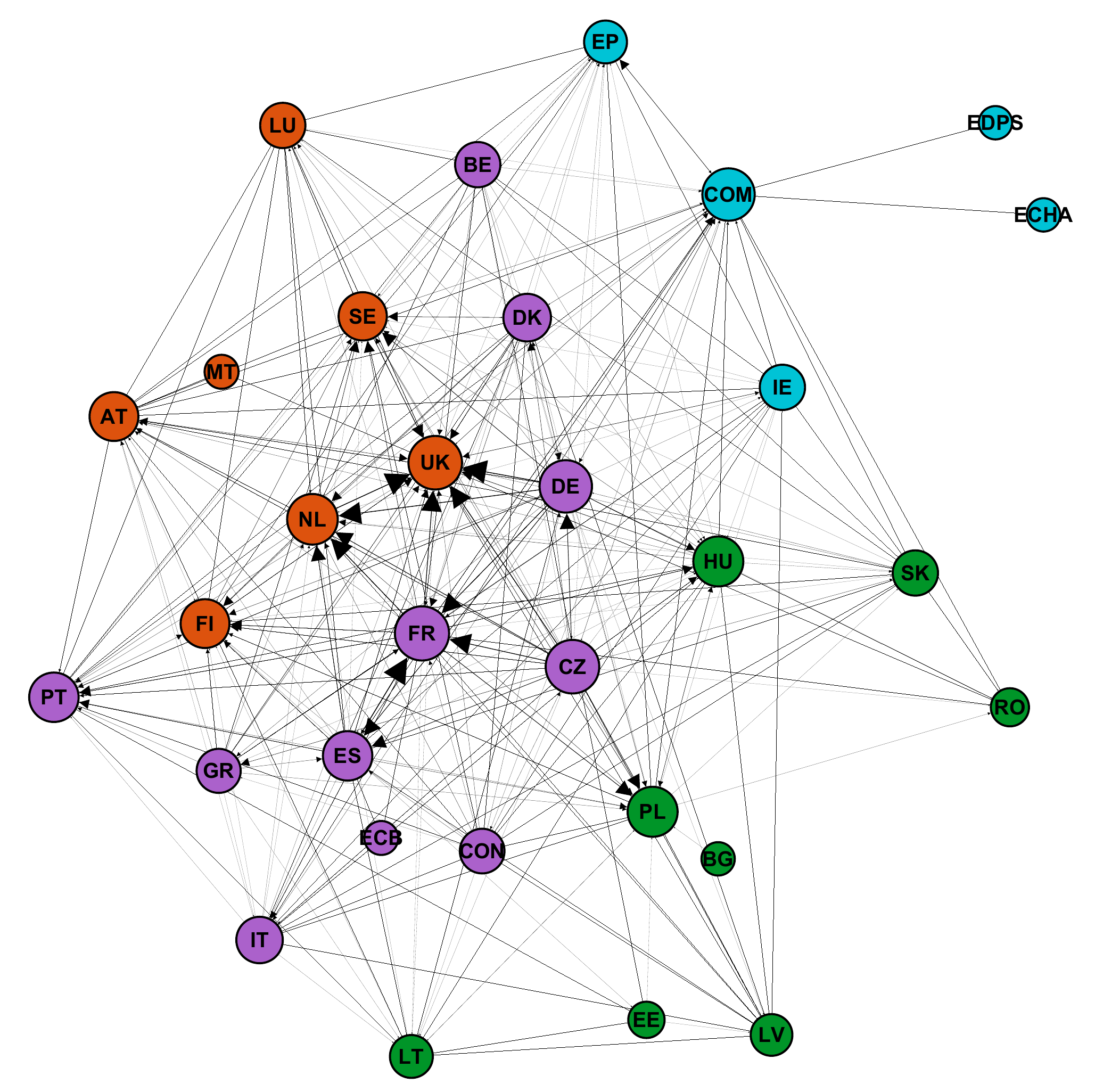}}
\hspace{5mm}
{\includegraphics[width=0.45\textwidth]{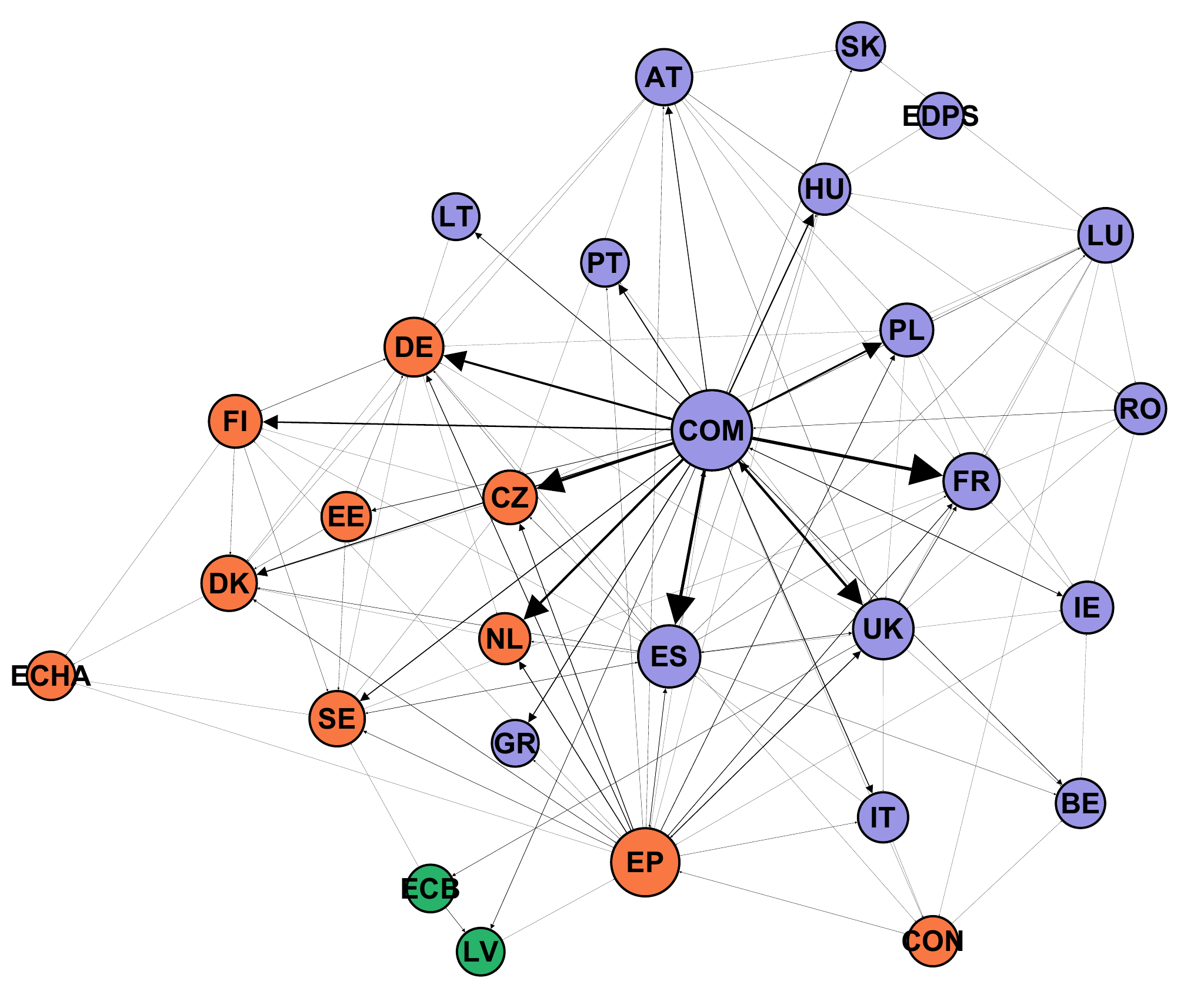}}
\caption{{\bf Network communities for 2012-2018.} Friends (left); Foes (right).\label{fig:comm}}
\end{figure}

\medskip
Fig.\ref{fig:comm} displays community structures, denoted by node colour, in the Friends and Foes networks for the last period under study (2012-2018). Communities in the Friends network reveal a clear separation between member states and pro-integration institutions countries as well as East-West and North-South divides, which social scientists have documented in legislative settings \cite{thomson2009actor}. The European Commission and the European Parliament, both in the light-blue community, tend to advocate federalist policies, whereas member states, along with the Council (which serves to represent the interests of national governments) typically advocate greater deference to domestic decision making \cite{ovadek2021supranationalism}. The purple community is mostly comprised of southern member states (France, Italy, Spain, Greece, Portugal) along with the Council. The orange community is a cluster of predominantly northern European member states (Sweden, Finland, Netherlands, UK). Eastern member states (including Poland, Hungary, Slovakia and Romania) form a separate (green) community. 

The position of the European Commission in the Foes network reflects its role as "guardian of the treaties". The Commission frequently intervenes to defend EU legislation against legal challenges brought by national governments. The European Parliament and the Council, who often defend opposite policy positions, find themselves in the same community (orange nodes). So do the UK, France and Spain -- a pattern largely driven by reciprocal hostility between the UK, on the one hand, and France and Spain, on the other.

\clearpage

\subsection{Multiplex Perspective}
\medskip
This section reports the results of our duplex analysis, which helps better understand the role of nodes as supporters/rivals. We focus on node degree, computing the overlapping degree between the two layers\cite{battiston2018multiplex} and the participation coefficient with respect to them\cite{avalos2018emergent}. The overlapping degree simply sums the node degrees over the two layers, while the participation coefficient quantifies the distribution of nodes presence in the two networks: it ranges in [0,1] and is equal to 0 if all edges of a node belong to just one layer, while it is exactly 1 if the edges are equally distributed over the two. 
\begin{figure}[!ht]
\centering
{\includegraphics[width=0.5\textwidth]{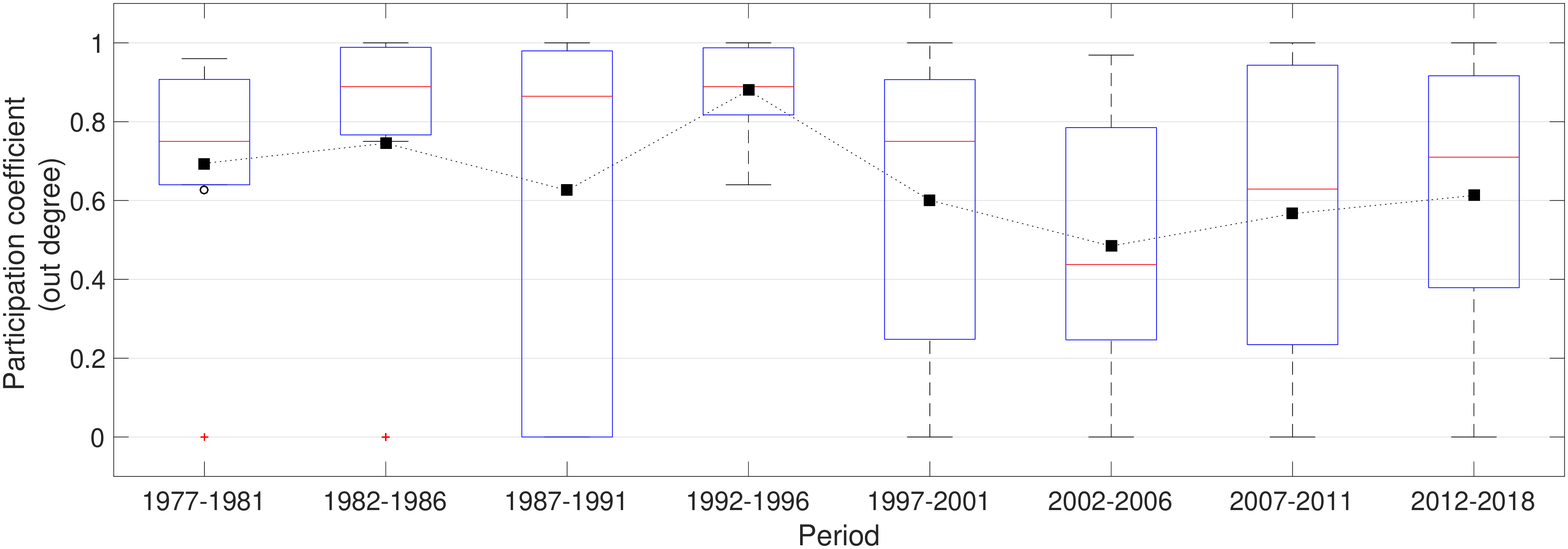}}
\hspace{-5mm}
{\includegraphics[width=0.5\textwidth]{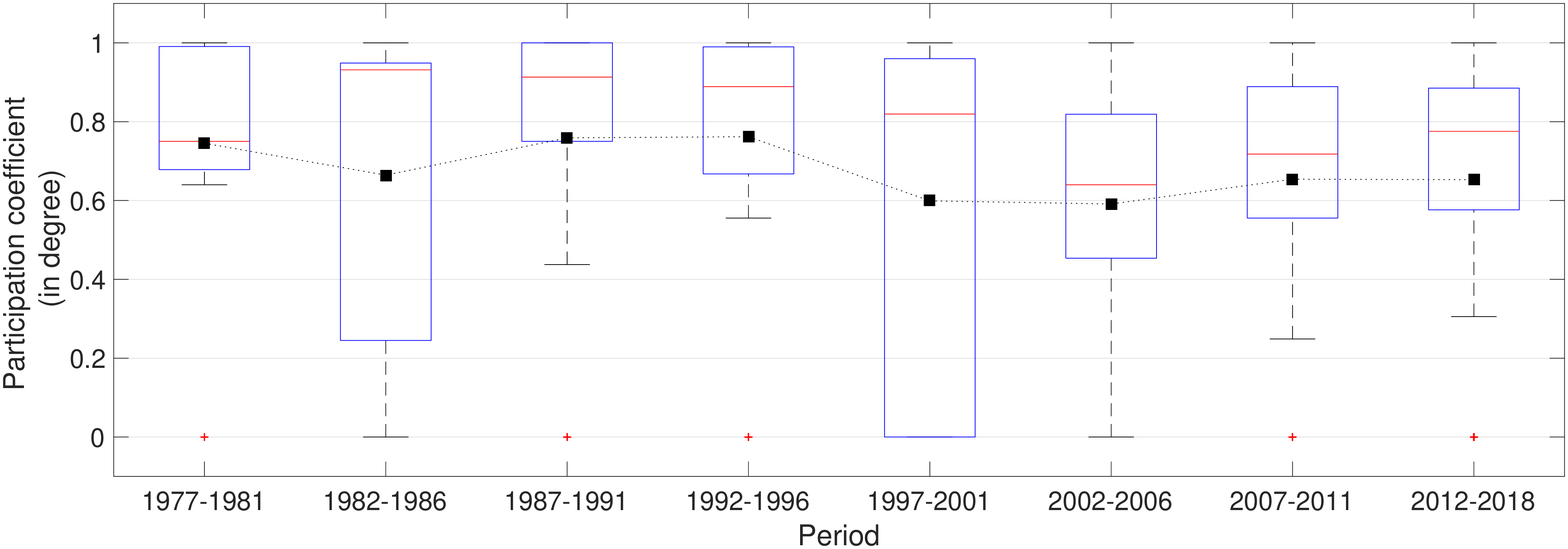}}
{\includegraphics[width=0.6\textwidth]{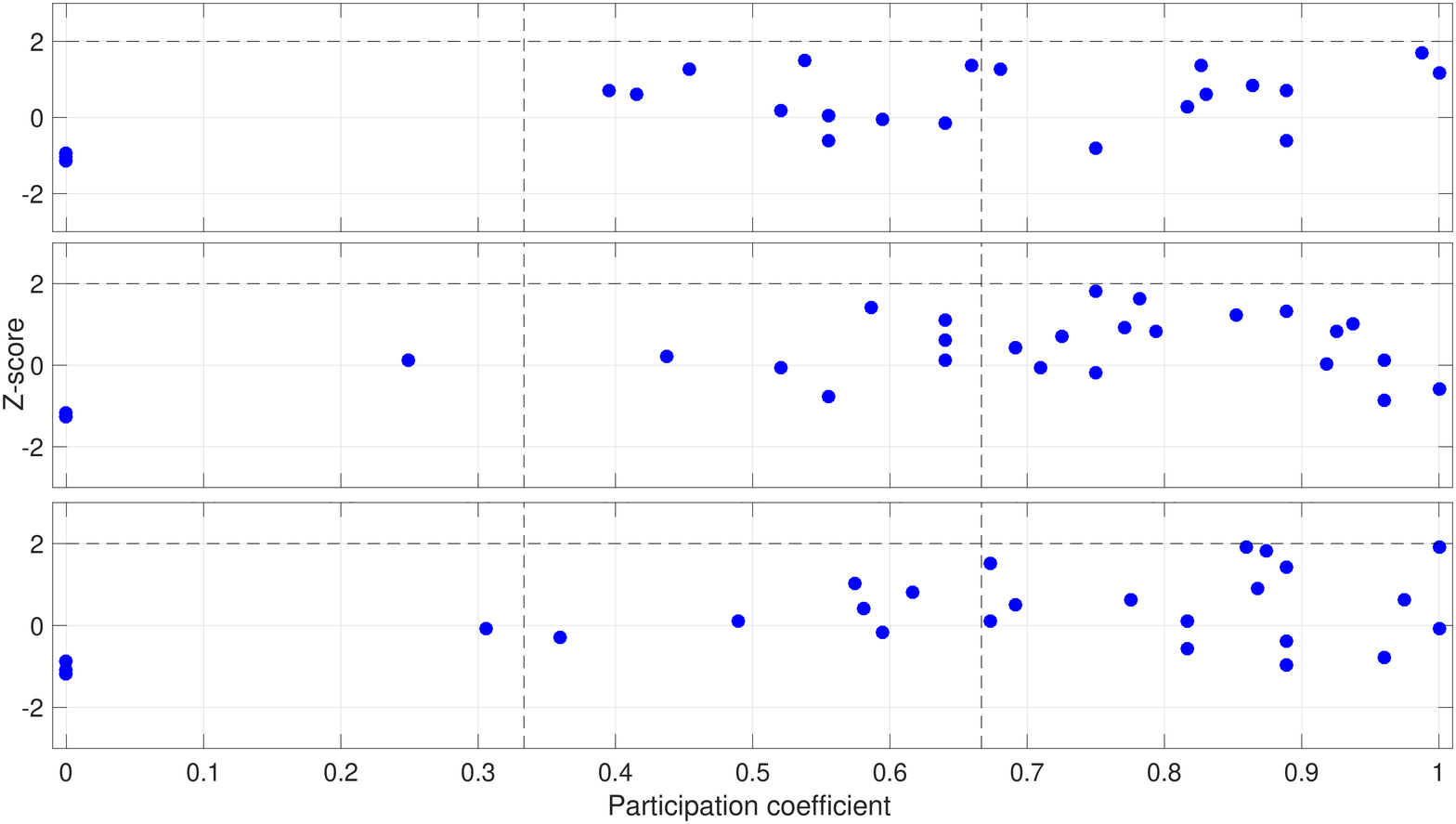}}
\hspace{13mm}
{\includegraphics[width=0.3\textwidth]{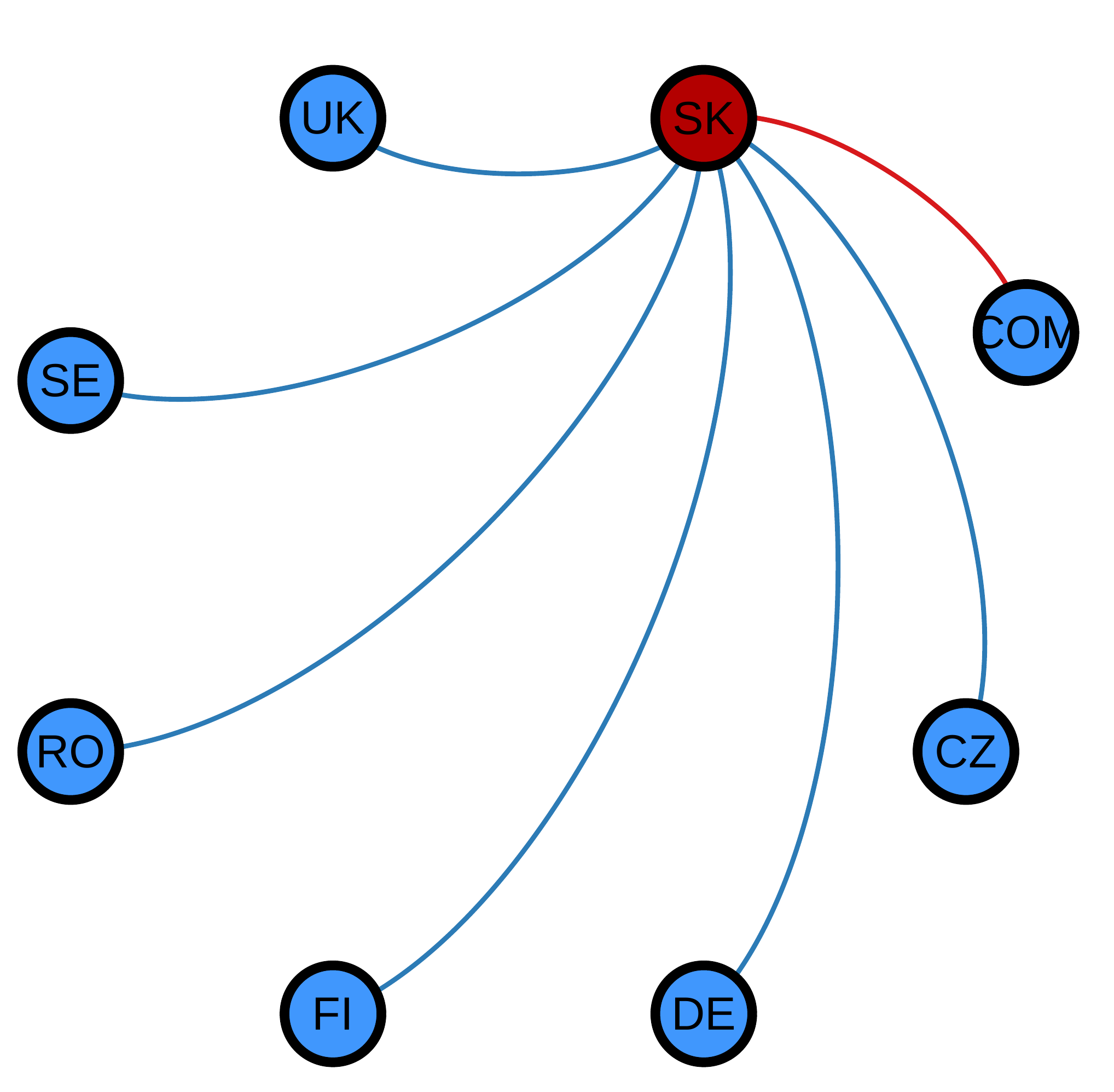}}
{\includegraphics[width=0.6\textwidth]{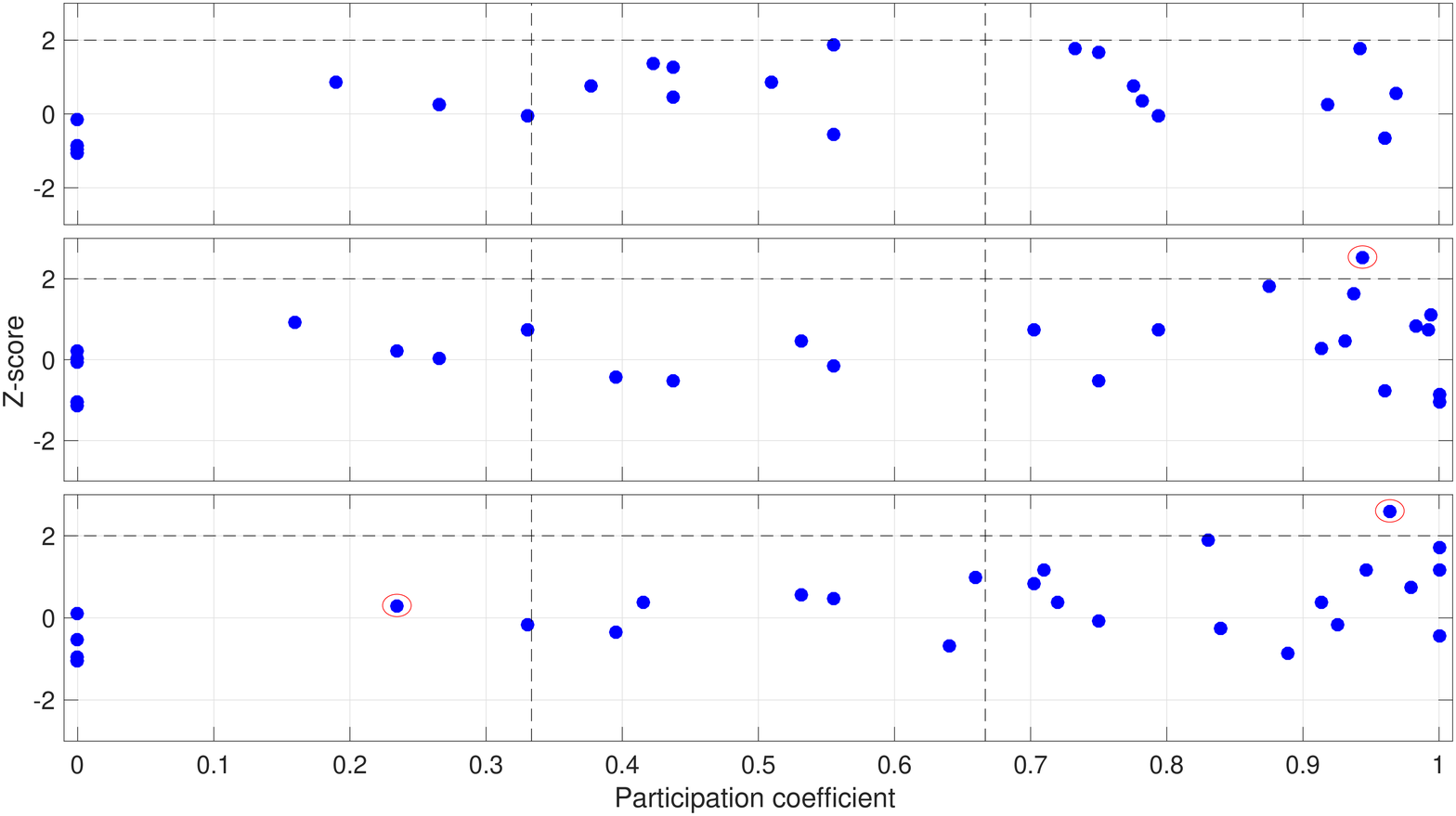}}
\hspace{13mm}
{\includegraphics[width=0.3\textwidth]{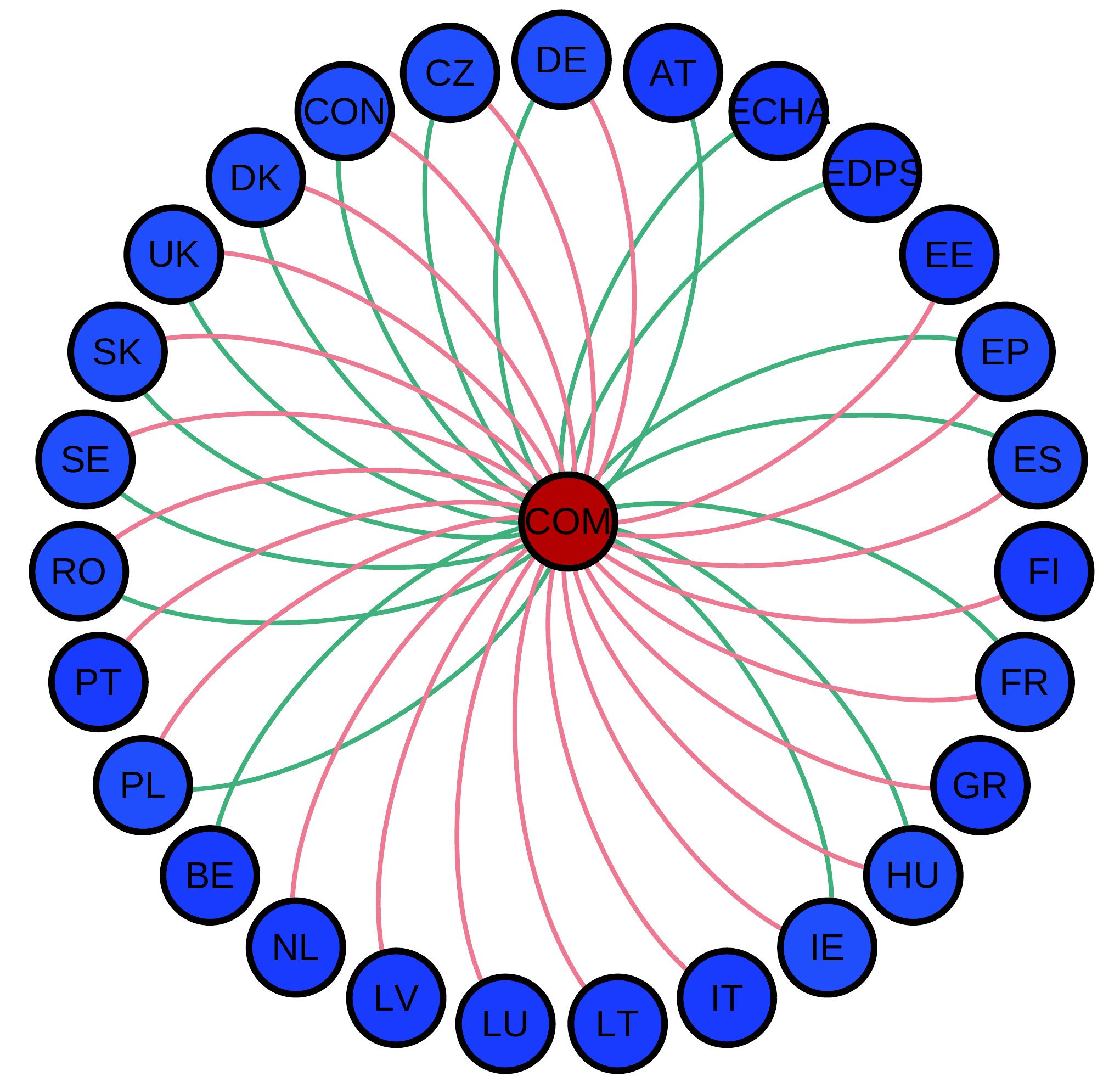}}
\caption{{\bf Role of nodes from a multiplex perspective.} Top: box plot of participation coefficient distribution for the 8 periods under study and the in-degree (left)/out-degree (right); Bottom left:the participation coefficient for the in-degree (top) and out-degree (bottom) versus the related z-score. We reported the results only for the last three periods under study. Bottom right: two examples from the out-degree case for the last period: the ego network of country SK and the institution COM (nodes red circled in the related plots).) \label{fig:rank}}
\end{figure}

A first inspection of node scores according to in/out overlapping degree reveals high variability in node behaviour, as shown in fig\ref{fig:rank}.
The same figure also reports the participation coefficient (with respect to in/out degree) versus the related z-scores, highlighting the role played by nodes in the two layers. Also displayed are the ego network of the Slovakia (SK) and the European Commission (COM). They were chosen to contrast the role of node in the duplex. Slovakia can be considered as a "focused" node, as its outgoing edges mainly belong to the Friends network (6, only one link in the other layer), whereas the European Commission exhibits a proper multiplex behaviour, behaving as hub in both layers with 16 outgoing links in the Friends and 25 in the Foes network.

\clearpage

\subsection{Merged Network}

\begin{figure}[!ht]
\centering
\subfigure[2002-2006]
{\includegraphics[width=0.45\textwidth]{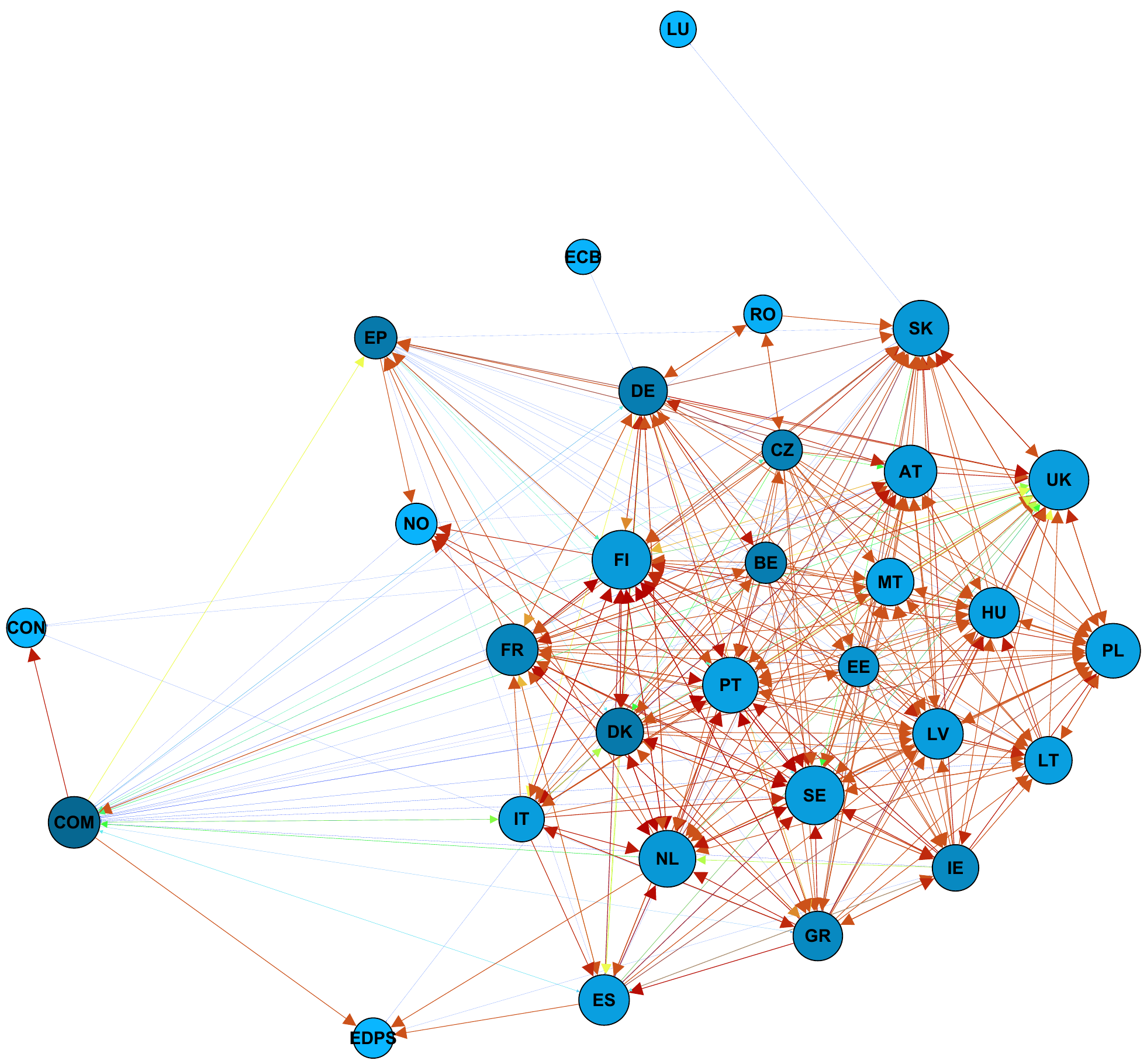}}
\subfigure[2007-2011]
{\includegraphics[width=0.45\textwidth]{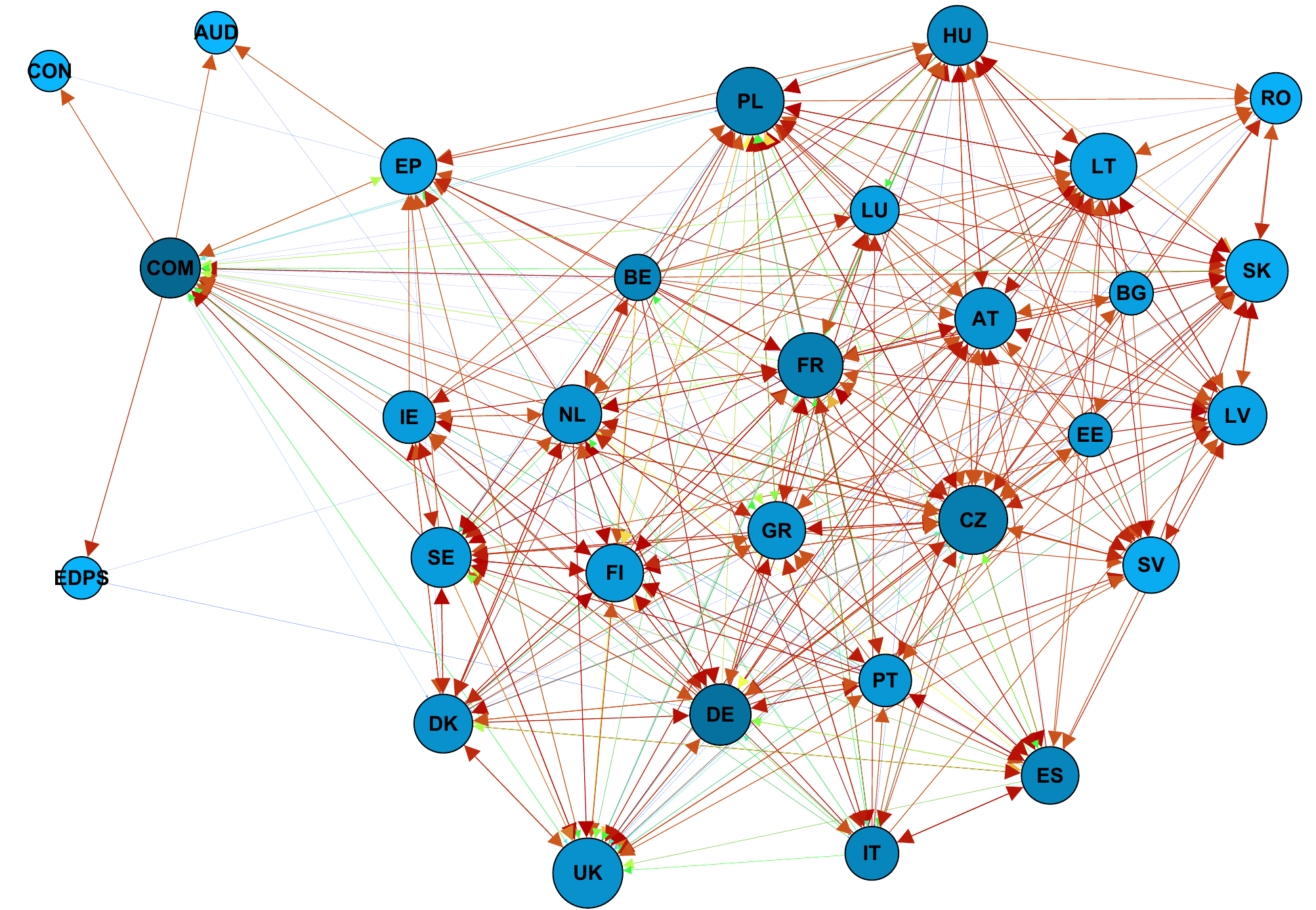}}
\subfigure[2012-2018]
{\includegraphics[width=0.45\textwidth]{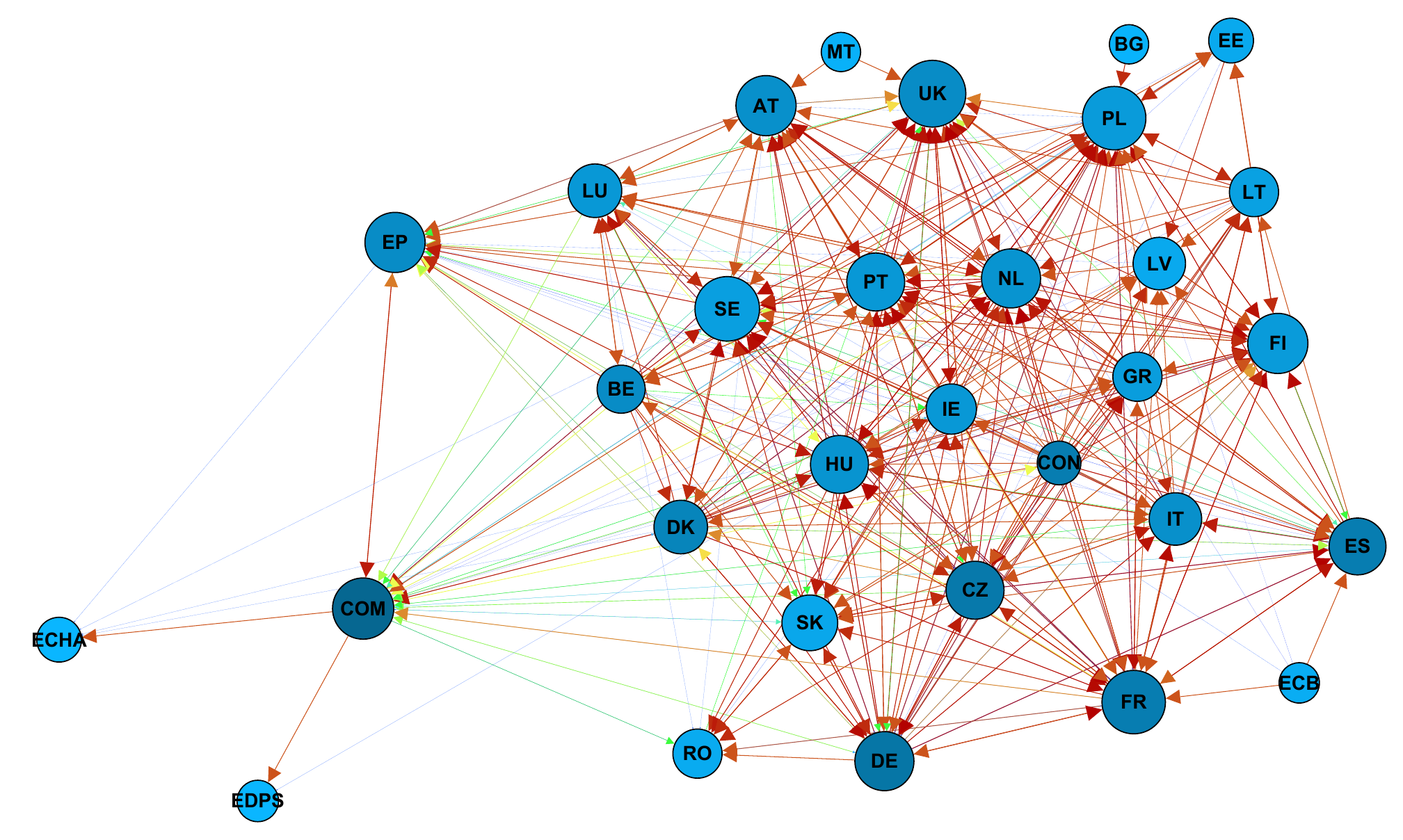}}
\caption{{\bf Networks of EU institutions and countries in litigation before the European Court of Justice.} Node size is directly proportional to in-degree; node colors is directly proportional to out-degree (from light to dark blue); edge color goes from blue to red proportionally to the number of cases involving the two nodes as supporters with respect to the total cases interesting them. \label{fig:new}}
\end{figure}
\medskip
In this section we define an alternative network merging information from both Friends and Foes connections among EU countries and institutions.

A link between any two nodes is weighted according to number of cases involving them as Friends divided by the total amount of cases between them:

\begin{equation}
\centering
W^{mrg} \equiv (w^{mrg}_{ij})_{1 \le i,j\le N}, \quad \text{with} \quad w^{mrg}_{ij} = w^{fr}_{ij}/(w^{fr}_{ij} + w^{fo}_{ij})
\end{equation}

The edge-direction indicates the source and the target of intervention. In other words, a link from A to B with weight 0.8 indicates that in 80\% of cases A supported B (regardless of whether A initiated the case or intervened later), while a weight equal to 0.1 means that only in 10\% of cases A supported B. 

Figure \ref{fig:new} illustrates the merged networks for the last three periods under study. Node size is proportional to in-degree centrality, while lighter shades of blue indicate lower and darker shades of blue higher out-degree centrality. Edge colour captures weights as defined above, with darker shades of red denoting more thoroughly supportive behaviour. The blue and green shades of the edges linking institutions and member states clearly show the divide opposing EU institutions and national governments in EU litigation. EU institutions typically promote pro-integration laws which national governments seek to contain \cite{ovadek2021supranationalism}.

\begin{figure}[!ht]
\centering
\subfigure[Out-going links]
{\includegraphics[width=0.9\textwidth]{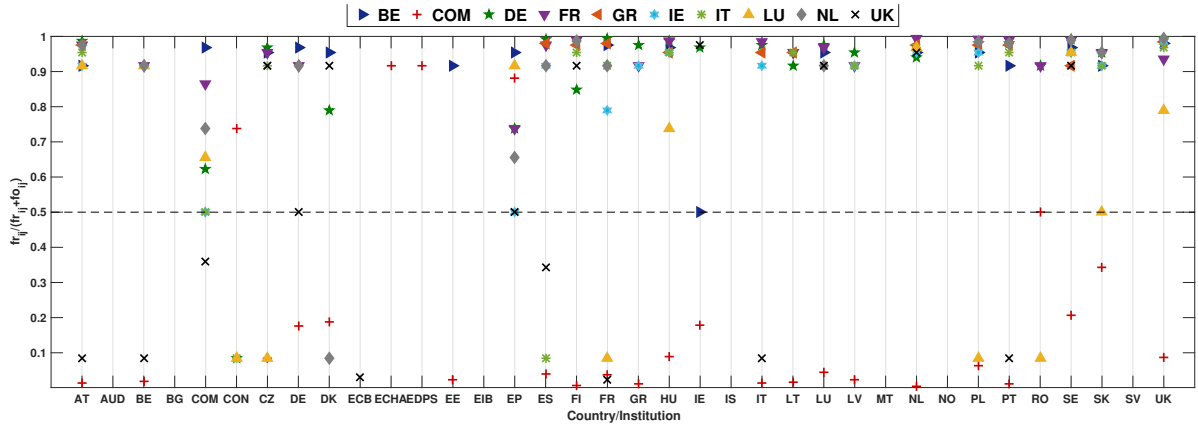}}
\subfigure[In-going links]
{\includegraphics[width=0.9\textwidth]{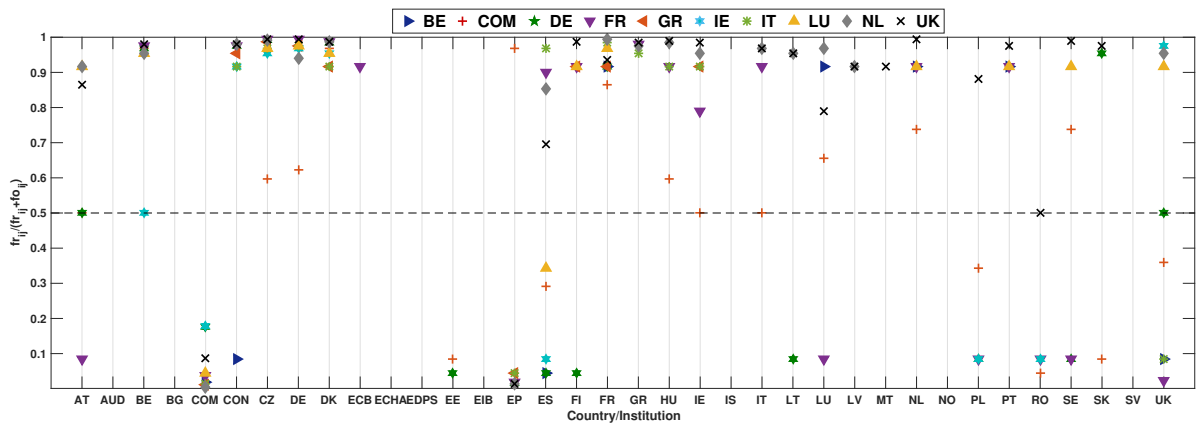}}
\caption{{\bf Frequency of supportive behaviour in 2012-2018.} Out-going (top) and in-going edge value of the novel network MRG: for each of the ten country/institution selected before we show the frequency of supporting behaviour towards a specific country/institution computed as $fr_{ij}/(fr_{ij}+fo_{ij})$ \label{fig:fre}}
\end{figure}

In figure \ref{fig:fre} we show for a select group of countries and institutions, how the frequency of supporting behaviour are distributed over all the country/institution present in the dataset. On the top we report out-going links, therefore the frequency associated to the country/institution indicated by the color as starter of the supportive action, while on the bottom we consider in-going link, therefore consider the node as target of the action started by the  country/institution indicated by the column.

Figure \ref{fig:fre} (a) shows interesting information about the tendency to be or not supportive in the last period under study. For example, COM appears generally not supportive as source of an intervention as most of exiting link values are small, except for connections with CON, ECHA, EDPS and EP, all European Institutions. The UK and Luxembourg direct supportive behaviour only towards some countries and institutions, respectively Czech Republic, Denmark, Finland,DK,FI,IE,LU,NL and SE for UK; EP, HU, NL,SE and UK for LU. 
Figure \ref{fig:fre} (b) reports information about receiving supportive interventions in the last period of our sample.. Most of the ten countries/institutions  reported here receive supportive behaviour from the others, with some exceptions for DE (small incoming edge links from COM, EE, ES, FI, LT), UK (small incoming edge links from COM and EP). These figures could help in attempting a classification for country/institution as mainly supportive or not, be supported or not and grouping agents accordingly. 



\section{Conclusions}

Our network analysis of third interventions before the CJEU provides insights into international litigation dynamics. The disassortative behaviour displayed in both networks indicates a tendency for nodes to form connections with dissimilar nodes rather than similar ones. The strong correlations among centrality measures suggest that certain member states and institutions hold a prominent role in litigation as source and target of interventions and in bridging the networks' communities. 
The modularity analysis revealed alignments along regional lines and divisions between EU institutions and member states, consistent with previous social science research on European integration. Lastly, the higher degree of reciprocity observed within the Foes network compared to the Friends network suggests a greater level of mutual opposition and conflict among nodes in the Foes network.

While our analysis remains exploratory, we hope to have showed that international litigation provides as a suitable context for network analysis, allowing researchers to navigate the complexity of the underlying coalition patterns.

\section*{Acknowledgements} 

RM acknowledges support from the Italian "Programma di Attivit\`a Integrata" (PAI) project "PROsociality COgnition and Peer Effects" (PRO.CO.P.E.), funded by IMT School for Advanced Studies Lucca and the European Union ? Horizon 2020 Program under the scheme ?INFRAIA-01-2018-2019 ? Integrating Activities for Advanced Communities?, Grant Agreement n.871042, ?SoBigData++: European Integrated Infrastructure for Social Mining and Big Data Analytics? (http://www.sobigdata.eu).

\section*{Author contributions}

RM, AD and GC conceived the study; RM analyzed the data, produced the figures and drafted the manuscript; RM, AD, GC interpreted the results. All authors critically revised the manuscript. 
\bibliography{lex}

\end{document}